\documentclass{article} %
\usepackage{iclr2027_conference,times}

\usepackage{amsmath,amsfonts,bm}

\def\eqref#1{equation~\ref{#1}}

\def\1{\bm{1}}

\DeclareMathAlphabet{\mathsfit}{\encodingdefault}{\sfdefault}{m}{sl}
\SetMathAlphabet{\mathsfit}{bold}{\encodingdefault}{\sfdefault}{bx}{n}

\usepackage{hyperref}
\usepackage{url}

\usepackage{subcaption}
\usepackage{natbib}
\usepackage{hyperref}
\usepackage{xcolor}
\definecolor{myblue}{HTML}{1a9ae8}  %
\hypersetup{
    colorlinks=true,
    linkcolor=myblue,
    citecolor=myblue,
    urlcolor=myblue,
    pdftitle={AudioTETRIS: Token Elimination via Transitional Importance Signals for Efficient Audio LLMs},
    pdfauthor={Shreya Sajid, Anna Vettoruzzo, Joaquin Vanschoren}
}
\usepackage{url}
\usepackage{graphicx}
\usepackage{setspace}
\usepackage{makecell}
\usepackage{booktabs}
\usepackage{colortbl}
\usepackage{wrapfig}
\usepackage{graphicx}
\usepackage{algorithm}
\usepackage{algpseudocode}

\usepackage{xcolor}
\usepackage{array}
\usepackage{multirow}
\usepackage{adjustbox}

\usepackage{threeparttable} %
\usepackage{xcolor}
\usepackage{tikz}
\usetikzlibrary{shapes.geometric}

\DeclareRobustCommand{\phonemedot}[1][blue!45]{%
  \tikz[baseline=-0.5ex]{
    \filldraw[fill=#1, draw=black, line width=0.3pt] (0,0) circle (2.2pt);
  }%
}

\DeclareRobustCommand{\graydot}{%
  \tikz[baseline=-0.5ex]{
    \filldraw[fill=gray!55, draw=none] (0,0) circle (2.2pt);
  }%
}
\DeclareRobustCommand{\starmark}[1][blue!45]{%
  \tikz[baseline=-0.6ex]{
    \node[star, star points=5, star point ratio=2.2,
          draw=black, fill=#1, line width=0.3pt,
          inner sep=0pt, minimum size=1.6ex] {};
  }%
}

\newcommand{\papertitle}{%
  \raisebox{-0.45\height}{\includegraphics[height=1.7cm]{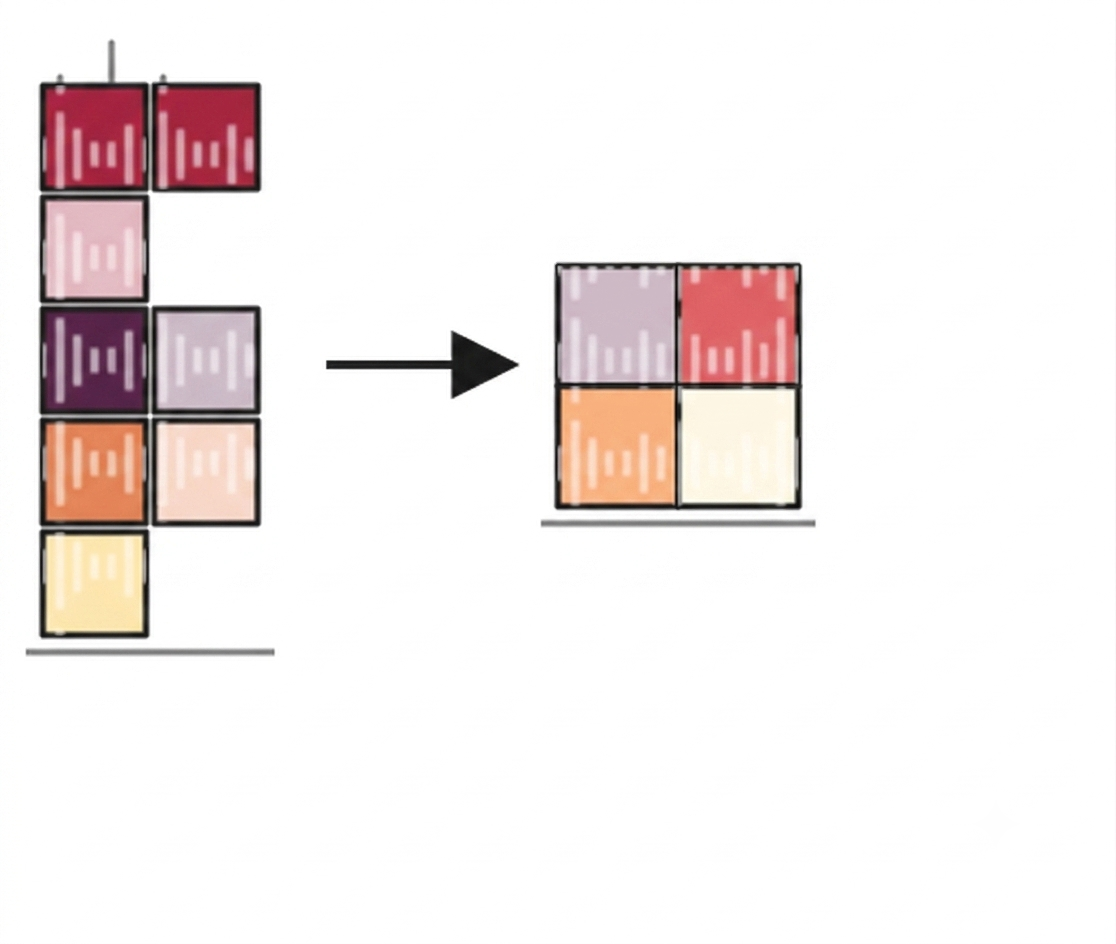}}%
  \hspace{-0.5em}%
   AudioTETRIS: Token Elimination via\\ Transitional Importance Signals for Efficient Audio LLMs
}

\title{\papertitle}

\author{Shreya Sajid, Anna Vettoruzzo \& Joaquin Vanschoren \\
Department of Mathematics \& Computer Science\\
Eindhoven University of Technology\\
Eindhoven, Netherlands \\
\texttt{\{s.sajid, a.vettoruzzo, j.vanschoren\}@tue.nl} 
}

\iclrfinalcopy
\begin{document}

\maketitle

\begin{abstract}
Large Audio Language Models (LALMs) achieve strong performance on speech-text tasks, but standard audio encoders produce hundreds of tokens for even short audio clips. Because self-attention scales quadratically in sequence length, this inflates inference-time compute substantially. Current training-free token compression methods prune or merge tokens based solely on coarse attention or feature similarity. However, treating audio tokens as generic, isotropic features smooths over spectral transitions, destroying fine-grained phonetic transitions essential for decoding speech. In contrast, speech signals exhibit highly structured temporal patterns where information density peaks at acoustic transitions. We propose \textbf{AudioTETRIS}, a training-free, two-stage framework that uses the audio encoder's representation geometry as an acoustic prior for token importance before applying attention-based compression. Instead of discarding encoder structure after the encoding stage, AudioTETRIS propagates this prior directly into the LLM decoder to compute a per-token fused score. This score drives progressive top-$k$ token selection across decoder layers without requiring additional training or auxiliary models. Evaluated on Qwen2-Audio-7B, Phi-4-Multimodal, and Audio Flamingo 3, AudioTETRIS retains near-uncompressed performance on several speech tasks using only 25\% of audio tokens, matches or surpasses full-token accuracy at moderate compression rates, and consistently outperforms existing training-free baselines.
\end{abstract}

\section{Introduction}
\label{sec:intro}

\begin{figure}[t]
    \centering
    \includegraphics[width=\columnwidth]{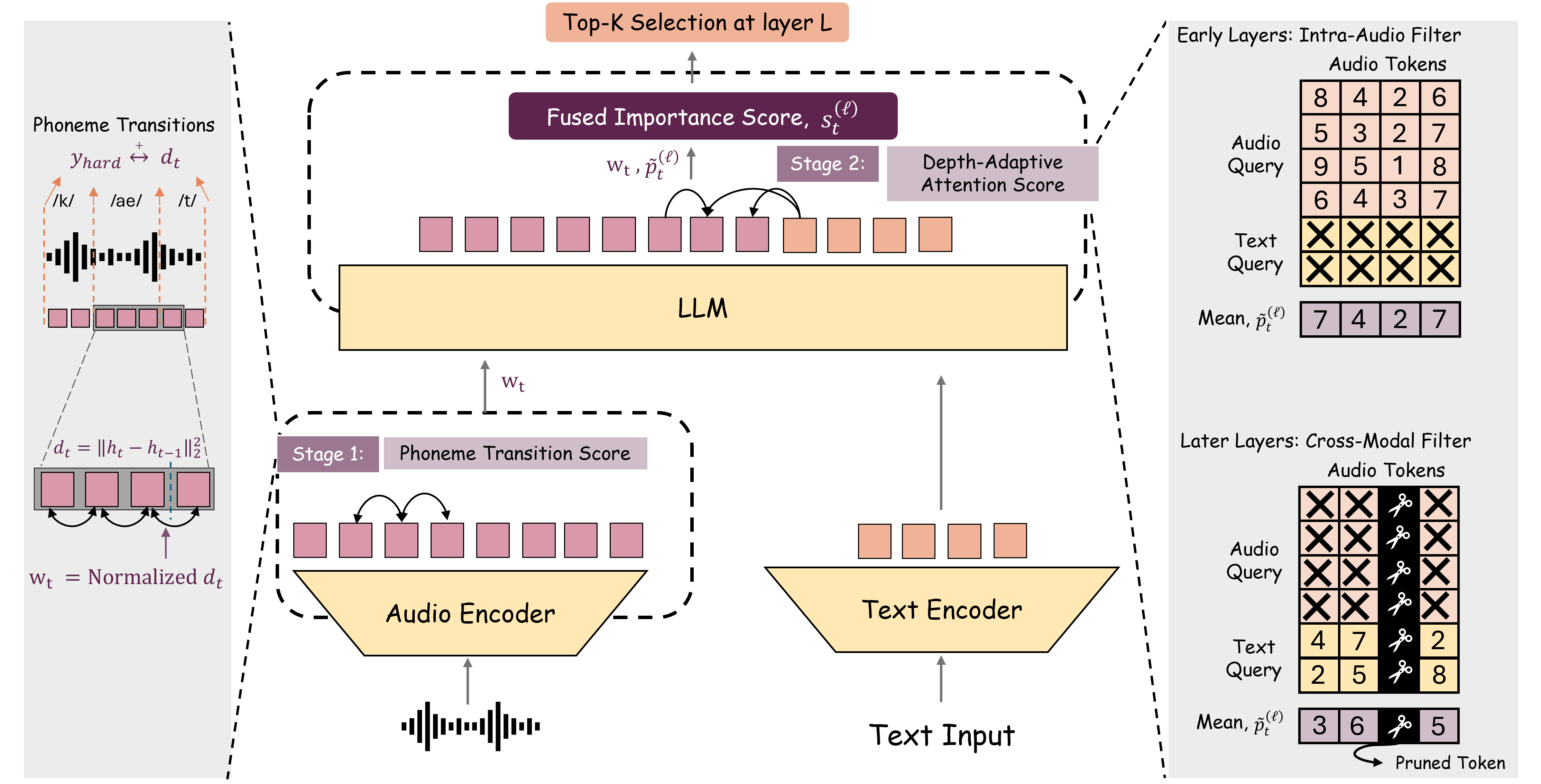}
    \caption{\textbf{Overview of AudioTETRIS.} A phoneme-transition-aware acoustic prior $w_t$ is computed from the audio encoder's hidden states via the transitional distance $d_t = \|h_t - h_{t-1}\|_2^2$, preserving phoneme transitions $y_{\text{hard}}$ (\S\ref{sec:analysis-prior}; left). Token pruning occurs in two stages: \textbf{Stage 1} injects $w_t$ as a static acoustic prior from the encoder's hidden states; \textbf{Stage 2} computes a depth-adaptive LLM attention score, fusing intra-audio attention in early layers with cross-modal (text-to-audio) attention in later layers (right). The two scores are fused into a single importance score, and top-$k$ selection retains the most salient tokens at each pruning layer $\ell$.}
    \label{fig:methodology}
    \vspace{-1.5em}
\end{figure}

Large Audio Language Models (LALMs) have achieved state-of-the-art results across several speech-text tasks like speech recognition, audio QA, and general audio understanding \citep{qwen2audio, whisper, QwenAudio2023, AudioBench2025, AIRBench2024}. Typically, these models use a pretrained audio encoder to map the input waveform into a sequence of continuous frame-level embeddings (audio tokens), which are projected into the Large Language Model (LLM) embedding space and prepended to text prompts. However, this approach creates a severe computational bottleneck: standard speech encoders generate hundreds of tokens for just a few seconds of audio \citep{speechprune, qwen2audio, goel2025audioflamingo3advancing}. Because decoder self-attention scales quadratically with sequence length, this large token footprint drives up inference costs \citep{10.1145/3530811}.

Furthermore, uniform audio tokenization is inherently redundant \citep{whisper, qwen2audio, goel2025audioflamingo3advancing, varstok, wertrap}. It clashes with the nature of speech, where information is concentrated during rapid phonetic transitions rather than steady-state regions \citep{furui1986role, stevens2002toward, aylett2004smooth, zhang2025distinctivefeaturecodecadaptive}. Reducing token count without losing these acoustically rich transition regions is vital for efficient LALM deployment. While recent methods apply token pruning or merging \citep{speechprune, bhati2025audio, echoingpixels, behera2024fastast}, they share a substantial flaw: they rely solely on isolated decoder attention or local feature similarity, ignoring the structural phoneme transitions that could be already captured by the encoder \citep{behera2024fastast, speechprune, echoingpixels, fastav2025, bhati2025audio}.

We hypothesize that speech token compression should be guided by the underlying acoustic structure of the signal, rather than post-hoc decoder attention alone. Our approach is based on a simple insight: regions of rapid spectral transition in the encoder's hidden states coincide with phonetic transitions, where essential acoustic details reside \citep{furui1986role, stevens2002toward}. We empirically validate this hypothesis in Section~\ref{sec:analysis-prior}, showing a strong correlation between representation variance in encoder space and actual phonetic segment transitions. Compression methods that ignore this structure could risk discarding high-frequency acoustic details and degrading downstream performance \citep{wertrap, dellalibera2026fixedframesdynamiccharacteraligned}.

To solve this, we introduce \textbf{AudioTETRIS} (Figure~\ref{fig:methodology}), a training-free framework that compresses audio tokens by jointly exploiting speech-specific acoustic structures and decoder-side attention. First, AudioTETRIS derives a phoneme-transition-aware prior from the temporal derivative of hidden states, isolating regions of high spectral change to protect information-dense transition segments. Second, this prior is propagated into the LLM decoder and fused with a depth-adaptive attention score derived from intra-audio attention in early layers and cross-modal text-to-audio attention in middle layers.  By coupling acoustic salience with task-conditioned attention, token survival is driven by both structural and semantic relevance.

We evaluate AudioTETRIS across backbones like Qwen2-Audio-7B, Phi-4-Multimodal, and Audio Flamingo 3 \citep{qwen2audio, microsoft2025phi4minitechnicalreportcompact, goel2025audioflamingo3advancing} on tasks including Automatic Speech Recognition (ASR), Speech Emotion Recognition, Audio Question Answering, and Speech Translation. AudioTETRIS retains near-lossless accuracy while keeping as few as 25\% of tokens, outperforming other training-free baselines, and even surpassing full-sequence models at moderate compression rates.

\newpage
Our main contributions are:
\begin{itemize}
    \item \textbf{Acoustic transition prior:} We show that temporal variance in encoder hidden states serves as a reliable proxy for phonetic transitions, capturing high-information regions without fine-tuning or external alignments.
    \item \textbf{AudioTETRIS framework:} We propose a training-free compression framework that propagates this acoustic prior into the LLM decoder, dynamically guiding self-attention to preserve both acoustic and semantic details.
    \item \textbf{State-of-the-art efficiency:} Extensive experiments demonstrate near-lossless performance at aggressive compression levels (25\% token retention), outperforming strong training-free baselines across diverse benchmarks.
\end{itemize}

\section{Related Work}
\label{sec:related}

\textbf{Large Audio Language Models.}
LALMs typically couple a pretrained speech encoder, such as Whisper~\citep{whisper}, with an LLM backbone~\citep{qwen2audio, audioflamingo2, goel2025audioflamingo3advancing, microsoft2025phi4minitechnicalreportcompact, tang2024salmonn, deshmukh2023pengi}, projecting audio features into the text embedding space. However, this paradigm yields audio token sequences that are orders of magnitude longer than the accompanying text, creating severe computational overhead during prefill and generation. Reducing this token count without sacrificing the dense lexical content of the underlying signal remains a key challenge \citep{shao2026survey, he2025audiomarathon}.

\textbf{Training-Free Token Compression.}
Training-free compression methods eliminate redundant audio tokens without modifying model weights. One direction seeks to prune or merge sequence tokens prior to or during LLM processing~\citep{bolya2023tome, behera2024fastast, chen2024fastv, fastav2025}. Across vision-language models (VLMs), methods like SparseVLM~\citep{zhang2025sparsevlmvisualtokensparsification} leverage self-attention logits to select text tokens as `raters' that adaptively prune visual tokens across layers~\citep[see also][]{Shang2025prumergeICCV, yang2025visionzip, xing2025pyramiddrop}. In the speech domain, SpeechPrune~\citep{speechprune} compresses long-form speech ($\ge$30s) tokens using approximated attention scores from the LLM’s initial transformer layer alongside speech-text feature similarity. However, relying on early-layer LLM dynamics forces the model to evaluate query relevance before building a comprehensive contextual representation of the signal. Alternatively, encoder-level methods like Local Temporal Bipartite Merging (LTBM)~\citep{ltbm} merge nearby tokens under a temporal window constraint~\citep{li23iATOMEinterspeech}, relying purely on local feature similarity rather than task-specific or phonetic cues.

\textbf{Training-Based \& Multimodal Token Compression.}
Rather than relying on training-free pruning, alternative strategies preserve performance by introducing extra trainable parameters or secondary modalities~\citep{kavaki2025audiosparse}. For instance, encoder-level pooling methods often require low-rank adapters (LoRA) \citep{hu2021loralowrankadaptationlarge} to fine-tune the backbone and recover downstream accuracy~\citep{bhati2025audio}. Although effective, this training dependency eliminates the module’s utility as a plug-and-play solution for unseen backbones or domain shifts. Other frameworks, such as EchoingPixels~\citep{echoingpixels}, achieve substantial token reduction by exploiting cross-modal alignment across visual and audio streams, but do not extend naturally to audio-language tasks. Across both directions, training-free methods bypass acoustic structure while structure-aware methods require fine-tuning or secondary modalities - a trade-off AudioTETRIS resolves with a training-free, acoustically grounded importance signal (\S\ref{sec:method}).

\section{Methodology}
\label{sec:method}

We present \textbf{AudioTETRIS}, a training-free framework for compressing audio tokens in LALMs. AudioTETRIS operates in two stages: Stage 1 (\S\ref{sec:method-prior}) extracts a phoneme-transition-aware acoustic prior directly from the audio encoder's hidden states, and Stage 2 (\S\ref{sec:method-attn}) propagates this prior into the LLM decoder, fusing it with a depth-adaptive attention score to progressively prune audio tokens across decoder layers. Figure~\ref{fig:methodology} illustrates the pipeline.

\subsection{Preliminaries}
\label{sec:method-prelim}

Given an input waveform, the audio encoder outputs a sequence of hidden states $\{h_1, \dots, h_T\} \in \mathbb{R}^{T \times d}$ representing $T$ audio tokens. These tokens are projected into the LLM embedding space and concatenated with the query embeddings to yield the initial decoder sequence $X = [x_1^a, \dots, x_T^a, x_1^q, \dots, x_M^q]$, with $a$ and $q$ referring to audio and query tokens. We index decoder layers as $\ell = 0, \dots, L_{\text{dec}}-1$, where $L_{\text{dec}}$ is the number of decoder layers, and denote the layer-$\ell$ self-attention weight matrix as $A^{(\ell)} \in \mathbb{R}^{H \times N^{(\ell)} \times N^{(\ell)}}$, where $H$ is the number of heads and $N^{(\ell)}$ is the sequence length at layer $\ell$ (which shrinks as audio tokens are pruned).

\subsection{Stage 1: Phoneme Transition Score}
\label{sec:method-prior}

\textbf{Motivation.} Linguistic information in speech is concentrated near acoustic-phonetic transitions~\citep{furui1986speaker}. Because these transitions carry the most acoustic and phonetic information, we hypothesize that tokens spanning them undergo significantly larger representational shifts in latent space than tokens within stable phonetic segments. We leverage this geometric property to construct an acoustic gate that identifies informative tokens without external supervision. 

\textbf{Phonetic Transition Score Formulation.} For each audio token $t$, we compute the token-to-token derivative $d_t$ as the squared Euclidean distance between consecutive hidden states $h_t, h_{t-1} \in \mathbb{R}^d$ taken from a single intermediate encoder layer (layer selection in \S\ref{sec:exp-setup}):
\begin{equation}
d_t = \| h_t - h_{t-1} \|_2^2, \qquad d_1 := 0,
\label{eq:dt}
\end{equation}
where the first token is assigned zero derivative in the absence of a preceding token. The squared $L_2$ norm amplifies sharp representational shifts over intra-segment noise at a linear computational cost of $\mathcal{O}(d)$ per token.

To modulate attention during pruning, we convert $d_t$ into a normalized acoustic weight $w_t$:
\begin{equation}
w_t = \frac{d_t}{\frac{1}{T}\sum_{i=1}^{T} d_i},
\label{eq:variance-weight}
\end{equation}
where $T$ is the total number of audio tokens in the utterance. The denominator enforces unit mean ($\mathbb{E}_t[w_t] = 1$).

\subsection{Stage 2: Depth-Adaptive Attention Scoring}
\label{sec:method-attn}

\begin{figure}[t]
    \centering
    \includegraphics[width=0.8\linewidth]{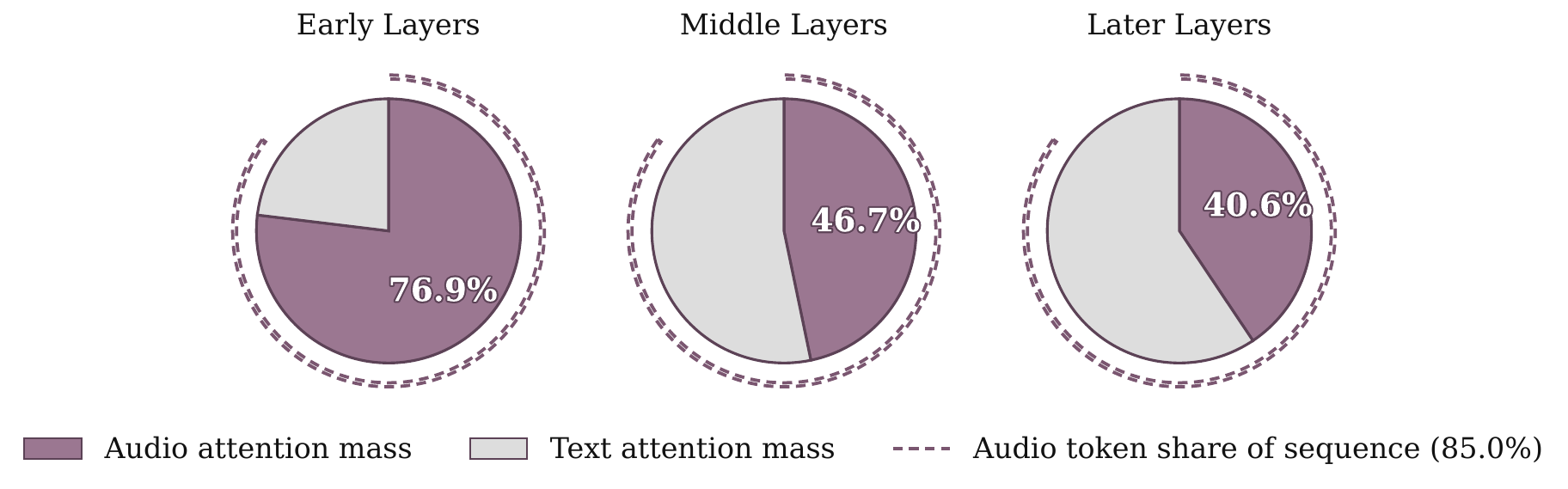}
    \caption{Distribution of decoder attention mass allocated to audio tokens across model depth in Qwen2-Audio on LibriSpeech. Despite audio tokens comprising $84.98\%$ of the input sequence (dashed ring), early layers are dominated by intra-audio attention ($76.91\%$), which drops sharply in middle ($46.73\%$) and late layers ($40.60\%$) as attention shifts to text-conditioned representations. This shift motivates our layer-dependent pooling rule in Eq.~\ref{eq:ptilde}.}
    \label{fig:attn_split_pies}
\end{figure}

\textbf{Motivation.} Attention mass allocation evolves significantly across LLM depth (Figure~\ref{fig:attn_split_pies}). Early decoder layers are dominated by intra-audio attention (audio tokens attending to audio tokens) to refine structural audio representations. In contrast, middle and late layers reallocate attention toward cross-modal interaction, where text queries attend directly to audio tokens. Consequently, a single, depth-invariant attention metric fails to capture token importance uniformly across the network.

\textbf{Layer-Adaptive Scoring.}
We address this depth dependence by defining the attention score $\tilde{p}_t^{(\ell)}$ for audio token $t$ at layer $\ell$ over a depth-specific subset of query rows:
\begin{equation}
\tilde{p}_t^{(\ell)} =
\begin{cases}
\dfrac{1}{H \cdot |\mathcal{Q}_a|} \displaystyle\sum_{h=1}^{H} \sum_{i \in \mathcal{Q}_a} A^{(\ell)}_{h, i, t}, & \ell \in \mathcal{L}_{\text{early}} \quad \text{(intra-audio filter)} \\[2.2ex]
\dfrac{1}{H \cdot |\mathcal{Q}_q|} \displaystyle\sum_{h=1}^{H} \sum_{i \in \mathcal{Q}_q} A^{(\ell)}_{h, i, t}, & \ell \in \mathcal{L}_{\text{later}} \quad \text{(cross-modal filter)}
\end{cases}
\label{eq:ptilde}
\end{equation}
where $\mathcal{Q}_a$ and $\mathcal{Q}_q$ denote the index sets of audio-query and text-query rows, $H$ is the number of attention heads, and $\{\mathcal{L}_{\text{early}}, \mathcal{L}_{\text{later}}\}$ partition the decoder layers.

\textbf{Automated Crossover Discovery.}
Rather than hand-crafting the layer partition $\{\mathcal{L}_{\text{early}}, \mathcal{L}_{\text{later}}\}$, we determine the crossover layer $\ell^{\dagger}$ from backbone attention dynamics on a small calibration set, as the first depth at which mean cross-modal attention mass $a_{\text{cross}}(\ell)$ overtakes mean intra-audio attention mass $a_{\text{intra}}(\ell)$:
\begin{equation}
\ell^{\dagger} = \min\{\ell : a_{\text{cross}}(\ell) \ge a_{\text{intra}}(\ell)\},
\label{eq:crossover-layer}
\end{equation}
This unsupervised procedure identifies $\ell^{\dagger}$ in a single calibration pass per backbone; full derivation and per-layer statistics are in Appendix~\ref{sec:appendix-crossover}.

\textbf{Score Fusion.} At each pruning layer $\ell$, we fuse the layer-adaptive attention score $\tilde p_t^{(\ell)}$ with the static acoustic gate $w_t$:
\begin{equation}
s_t^{(\ell)} = \tilde p_t^{(\ell)} \cdot w_t.
\label{eq:fusion}
\end{equation}
At each pruning layer $\ell$, we rank tokens by $s_t^{(\ell)}$ and retain a top fraction $k_\ell$, discarding the rest (see \S\ref{sec:exp-setup} for the pruning schedule and how $k_\ell$ is set per compression rate).

\section{Analysis: Validation of Acoustic State Dynamics}
\label{sec:analysis-prior}

In Section~\ref{sec:method-prior}, we hypothesized that token-to-token encoder state transitions $d_t$ track actual phonetic transitions. Here, we validate this hypothesis both quantitatively against ground-truth forced alignments and label-free via distributional kurtosis.

\textbf{Phoneme Transition Correlation.}
To evaluate whether encoder distance spikes align with physical phoneme transitions, we define a continuous diagnostic transition prior $b_t$. Because audio encoders distribute context across neighboring tokens~\citep{whisper}, we convolve $d_t$ with a Gaussian kernel of width $\sigma = 2$ and clip to $[0, 1]$:
\begin{equation}
b_t = \mathrm{clip}\Big(\textstyle\sum_{\tau} d_\tau \cdot \mathcal{N}(t - \tau;\, 0,\, \sigma^2),\ 0,\ 1 \Big).
\label{eq:bt}
\end{equation}

We evaluate $b_t$ against forced alignments on  LibriSpeech~\citep{panayotov2015librispeech, librispeech_alignments_github}. Standard alignment tools yield discrete phonetic transitions $y_{\text{hard}}$. To model continuous speech coarticulation~\citep{furui1986speaker}, we apply identical Gaussian smoothing to $y_{\text{hard}}$, producing a continuous target $y_{\text{soft}}$. We measure the Pearson correlations $r_{\text{hard}} = \mathrm{corr}(b_t, y_{\text{hard}})$ and $r_{\text{soft}} = \mathrm{corr}(b_t, y_{\text{soft}})$. Figure~\ref{fig:boundary-schematic} in Appendix~\ref{sec:appendix-boundary-schematic} illustrates this end-to-end pipeline. We provide additional dataset validations in Appendix~\ref{sec:appendix-covost-iemocap}. 

\begin{figure}[t]
    \centering
    \includegraphics[width=0.85\textwidth]{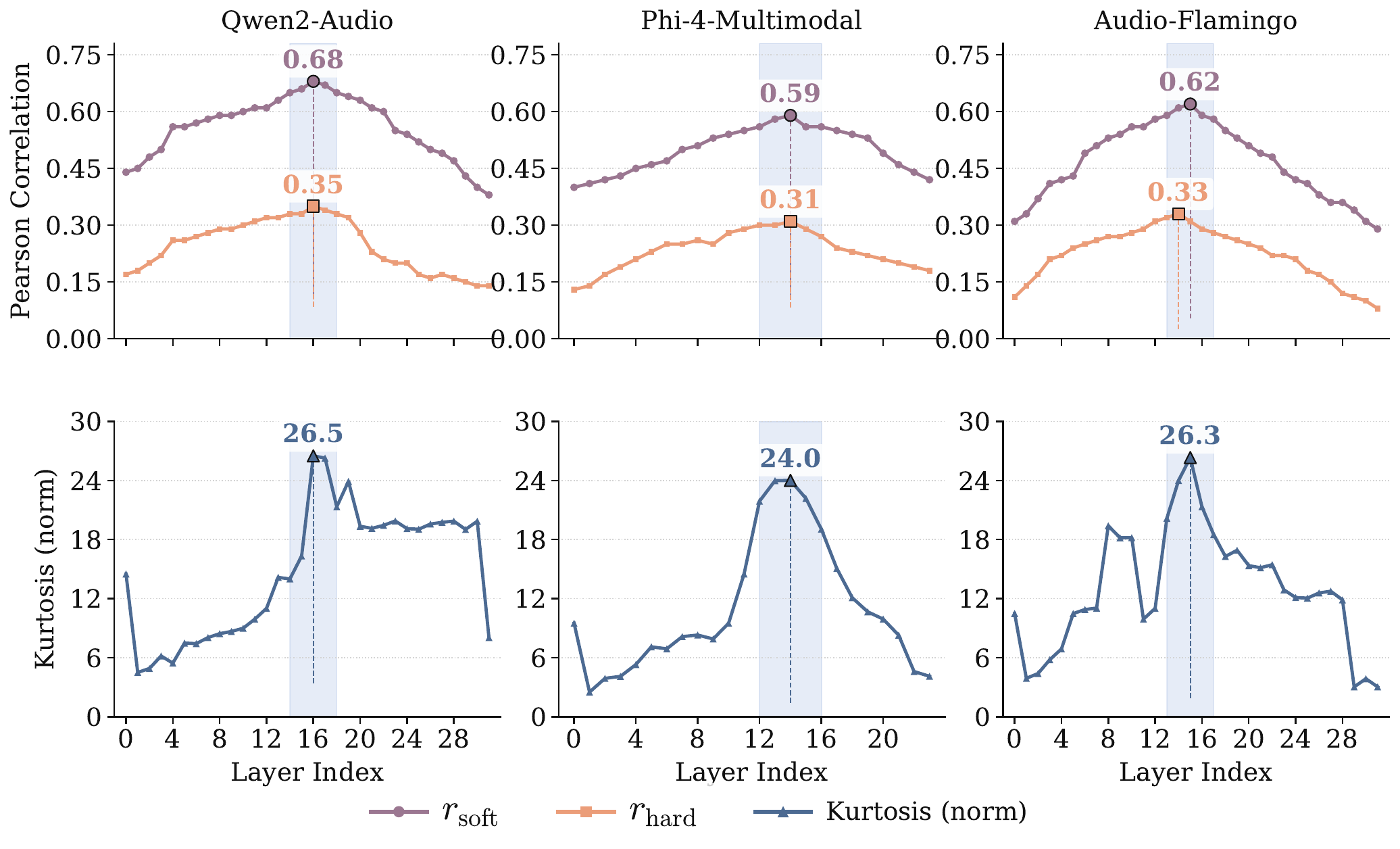}
    \caption{
        Layer-wise diagnostics across Qwen2-Audio, Phi-4-Multimodal, and Audio Flamingo 3. \textbf{Top:} Mean correlation between $b_t$ and ground-truth phoneme transitions ($r_{\text{soft}}$ and $r_{\text{hard}}$). \textbf{Bottom:} Kurtosis ($\kappa$) of the distance distribution $d_t$. Both metrics identify an intrinsic mid-network sweet spot (shaded) where phonetic transitions are maximally accentuated.
    }
    \label{fig:layer_correlations}
\end{figure}

As shown in Figure~\ref{fig:layer_correlations} (top), transition alignment varies predictably across depth, peaking in mid-network layers across all backbones: Qwen2-Audio (layer 16, $r_{\text{soft}} = 0.68$) \citep{qwen2audio}, Phi-4-Multimodal (layer 14, $r_{\text{soft}} = 0.59$) \citep{microsoft2025phi4minitechnicalreportcompact}, and Audio Flamingo 3 (layer 15, $r_{\text{soft}} = 0.62$) \citep{goel2025audioflamingo3advancing}. Appendix~\ref{sec:appendix_sigma_ablation} provides an ablation over $\sigma$.

\textbf{Label-Free Diagnostic via Kurtosis.}
To verify this mid-network correlation peak without relying on external alignments, we measure the distributional kurtosis ($\kappa$) of $d_t$ across layer depth. High kurtosis indicates that transitions stand out as sharp, heavy-tailed outliers above background intra-segment noise. As shown in Figure~\ref{fig:layer_correlations} (bottom), kurtosis peaks at the exact same mid-network layers ($\kappa \approx 24.0$-$26.5$), providing an unsupervised method to verify transition isolation.

\textbf{Trajectory Visualization.} A qualitative t-SNE analysis of per-token hidden states (Appendix~\ref{section:appendix-trajectory}, Figure~\ref{fig:embedding_trajectory}) corroborates this pattern: mid-network layers show tight phoneme clustering separated by sudden trajectory jumps at transitions, while early and late layers show progressively less phonetically consolidated structure.

\section{Experiments \& Results }

\subsection{Experimental Setup}
\label{sec:exp-setup}

\textbf{Backbones and Datasets.} We evaluate AudioTETRIS on three LALM backbones with distinct encoder-decoder architectures: Qwen2-Audio-7B ($\sim$8.4B)~\citep{qwen2audio}, Phi-4-Multimodal (5.6B)~\citep{microsoft2025phi4minitechnicalreportcompact}, and Audio Flamingo 3 ($\sim$8.3B)~\citep{goel2025audioflamingo3advancing}. Our evaluation spans four core tasks: automatic speech recognition on LibriSpeech test-other~\citep{panayotov2015librispeech}, emotion recognition on IEMOCAP \citep{iemocap}, audio question answering on MMAU-mini~\citep{MMAU2024}, and speech translation (English-to-Chinese) on CoVoST 2~\citep{wang2020covost2massivelymultilingual}. Together, these benchmarks span lexical transcription, cross-lingual generation, paralinguistic classification, and complex audio reasoning, allowing us to test whether preserving phoneme-transition regions specifically benefits tasks that require precise lexical recovery.

\textbf{Baselines.} We compare against SpeechPrune~\citep{speechprune}, SparseVLM~\citep{zhang2025sparsevlmvisualtokensparsification}, and LTBM~\citep{ltbm}, representing the two families of prior work discussed in \S\ref{sec:related}. We additionally include a Random Pruning baseline that retains a uniformly random token subset at matched compression rates, providing a lower-bound sanity check. All baselines are implemented under matched compression schedules using released hyperparameters where available.

\textbf{Implementation Details.} We evaluate four target token compression rates - $0.36, 0.51, 0.64$, and $0.75$ - representing the fraction of audio tokens removed by the final decoder layer. These correspond to uniform per-stage retention factors $r \in \{0.8, 0.7, 0.6, 0.5\}$ across two pruning layers ($1 - r^2$) in our experiments, ranging from light reduction to aggressive $4\times$ compression. Pruning is scheduled at layer~1\footnote{Decoder layers are indexed starting from 0 throughout this paper (matching the axis convention used in Figures~\ref{fig:attn_split_pies} and~\ref{fig:crossover_calibration}); e.g., `layer 1' denotes the second decoder layer.} and at a backbone-specific later layer
$\ell^{\dagger}$ (Eq.~\ref{eq:crossover-layer}) via progressive top-$k$ retention on the fused score $s_t^{(\ell)}$  (Eq.~\ref{eq:fusion}). The acoustic prior $w_t$ is computed from a single mid-network encoder layer per backbone: layer 16 for Qwen2-Audio, layer 14 for Phi-4-Multimodal, and layer 15 for Audio Flamingo 3. These layers are selected by the label-free kurtosis criterion of \S\ref{sec:analysis-prior} (the depth maximizing $\kappa$ of $d_t$); forced alignments are used only to validate this choice post hoc and are not required to apply AudioTETRIS to a new backbone. The diagnostic prior $b_t$ is computed from the same layer. All experiments are executed on a single NVIDIA A100 GPU. Layer choices, calibration, and alternative schedules are detailed in Appendices~\ref{app:layer-selection} and~\ref{app:multi-stage}.
 
\newpage
\subsection{Main Results}
\label{sec:exp-main}

We report overall performance and efficiency across four compression rates on Qwen2-Audio in Table~\ref{tab:main_results}. Three key observations emerge from these results:

\textbf{AudioTETRIS maintains high accuracy even under heavy compression.} At low compression (rate 0.36), AudioTETRIS matches or improves upon the uncompressed baseline on many benchmarks (4.52 WER, 58.04\% IEMOCAP, 42.36\% MMAU-mini vs.\ 4.61, 55.11\%, 42.20\% uncompressed). Even at rate 0.75, WER stays at 8.50 while IEMOCAP, MMAU-mini, and CoVoST2 remain within 2 points of uncompressed baselines, confirming that our method removes redundant tokens without disrupting essential lexical and semantic transitions.

\textbf{Performance gaps widen as compression grows aggressive.} At rate 0.36, several baselines remain competitive; by rate 0.75, baseline methods degrade sharply (random pruning WER climbs to 31.59), lacking a mechanism to safeguard acoustic transitions. AudioTETRIS outperforms the best baseline by over 1.2 WER points at the highest compression setting.

\textbf{ASR is inherently more vulnerable to token pruning.} Fine-grained transcription (WER) degrades far faster under aggressive compression than high-level understanding (MMAU-mini) or emotion recognition (IEMOCAP): at rate 0.75, IEMOCAP and MMAU-mini stay nearly intact while WER roughly doubles. High-level tasks rely on coarse acoustic features and tolerate token dropout, whereas exact lexical recovery needs fine-grained transition preservation - a vulnerability that speech-specific methods (AudioTETRIS, SpeechPrune) mitigate far better than general-purpose vision baselines or non-speech heuristics.

\subsection{Efficiency Analysis}
\label{sec:exp-efficiency}
Pruning within the LLM backbone yields direct computational savings, dropping total TFLOPs from 7.89 to 4.32 at a 0.75 compression rate, with latency falling from 1.31\,s to 0.62\,s over the same range. In-LM methods (AudioTETRIS, SparseVLM) incur slightly higher TFLOPs than single-shot methods (SpeechPrune, LTBM) since early layers still process the full, unpruned sequence before the first pruning stage; this difference is minimal relative to AudioTETRIS's accuracy gains. Real-world latency scales somewhat less than the FLOP drop due to RoPE reindexing and KV-cache trimming overhead at intermediate layers (Appendix~\ref{app:attn-backend}).

\subsection{Generalization Across Backbones}
\label{sec:exp-generalization}

\begin{figure}[t]
    \centering
    \includegraphics[width=\linewidth]{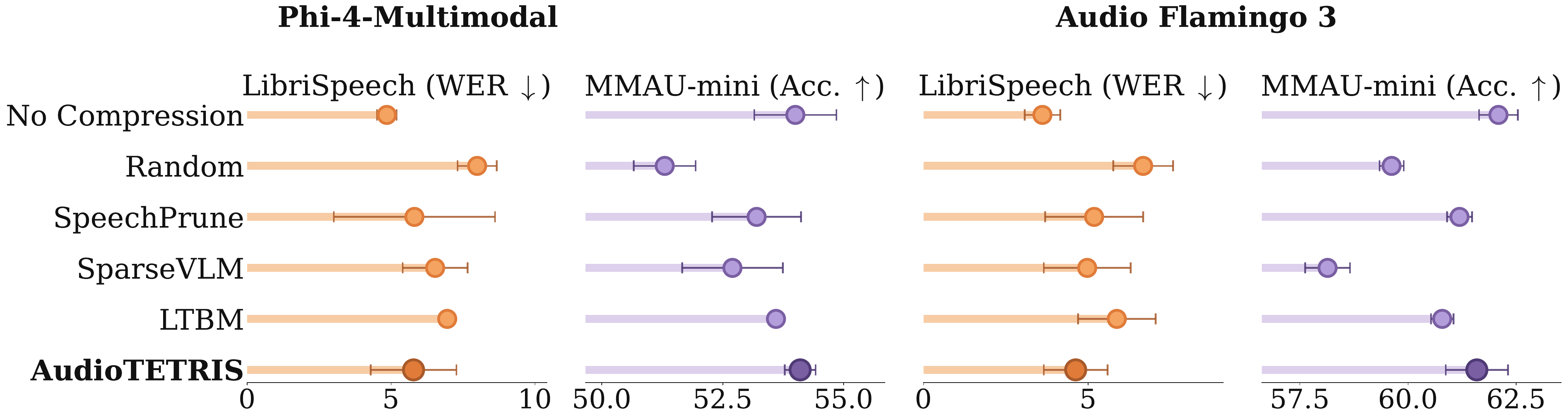}
    \caption{Comparison of LibriSpeech WER ($\downarrow$) and MMAU-mini accuracy ($\uparrow$) at compression ratio 0.51, for Phi-4-Multimodal and Audio Flamingo~3. Error bars show standard deviation across runs. \textsc{AudioTETRIS} is highlighted in the darker shade.}
    \label{fig:wer_mmau_lollipop}
\end{figure}

To test whether AudioTETRIS's gains are specific to Qwen2-Audio's architecture, we repeat the main comparison on Phi-4-Multimodal and Audio Flamingo 3 at compression ratio $0.51$ (Figure~\ref{fig:wer_mmau_lollipop}). On both backbones, AudioTETRIS remains within 1 point of uncompressed WER and attains the strongest or near-strongest accuracy on MMAU-mini, while baselines that match it on one metric (e.g., WER) fall sharply behind on others (e.g., SpeechPrune loses over 17 BLEU points on CoVoST2); Full numbers are given in Table~\ref{tab:phi4_appendix} (Phi-4-Multimodal) and Table~\ref{tab:audioflamingo3_appendix} (Audio Flamingo 3).

These results suggest the phonetic transition prior transfers across encoders with different subsampling rates and architectures (Phi-4-Multimodal's Conformer encoder vs.\ the Whisper-based encoders of Qwen2-Audio and Audio Flamingo 3), consistent with encoder-state transitions being a general property of speech representations rather than an artifact of one backbone. Extended results across all four compression ratios and benchmarks are in Appendix~\ref{sec:appendix_multibackbone}, where similar patterns hold.

\definecolor{highlightorange}{HTML}{F8DACD}
\begin{table}[t]
\centering
\caption{Performance and efficiency comparison across compression ratios on Qwen2Audio evaluated on LibriSpeech (ASR), IEMOCAP (Emotion Recognition), MMAU-mini (Audio Understanding, MCQ), and CoVoST2 (Speech Translation). WER (\%) is lower-is-better ($\downarrow$); Accuracy (\%) and BLEU are higher-is-better ($\uparrow$). Baselines: SpeechPrune~\citep{speechprune}, SparseVLM~\citep{zhang2025sparsevlmvisualtokensparsification}, LTBM~\citep{ltbm}.}
\label{tab:main_results}
\begin{threeparttable}
\setlength{\tabcolsep}{4pt}
\resizebox{\textwidth}{!}{%
\begin{tabular}{@{}lcccccc@{}}
\toprule
\textbf{Method} & \textbf{LibriSpeech WER} $\downarrow$ & \textbf{IEMOCAP Acc.} $\uparrow$ & \textbf{MMAU-mini Acc.} $\uparrow$ & \textbf{CoVoST2 BLEU} $\uparrow$ & \textbf{Latency (s)} & \textbf{TFLOPs} \\
\midrule
No Compression & $4.61^{\pm 0.12}$ & $55.11^{\pm 1.06}$ & $42.20^{\pm 0.96}$ & $33.03^{\pm 0.72}$ & $1.31$ & $7.89$ \\
\midrule
\multicolumn{7}{c}{\textit{Compression Ratio 0.36}} \\
\midrule
Random & $9.91^{\pm 1.67}$  & $53.10^{\pm 0.37}$ & $40.18^{\pm 0.23}$ & $29.10^{\pm 0.02}$ & $1.15$ & $6.93$ \\
SpeechPrune & $6.75^{\pm 1.44}$  & $56.43^{\pm 0.35}$ & $41.17^{\pm 0.60}$ & $31.34^{\pm 0.07}$ & $1.21$ & $6.99$ \\
SparseVLM$^{\dagger\ddagger}$ & $8.40^{\pm 1.08}$  & $52.37^{\pm 0.21}$ & $41.09^{\pm 0.73}$ & $31.38^{\pm 0.12}$ & $1.16$ & $7.28$ \\
LTBM & $5.90^{\pm 0.48}$  & $55.90^{\pm 0.27}$ & $40.67^{\pm 0.67}$ & $29.97^{\pm 0.06}$ & $1.25$ & $6.93$ \\
\textbf{AudioTETRIS}$^{\dagger}$ & \cellcolor{highlightorange}$4.52^{\pm 0.21}$ & \cellcolor{highlightorange}$58.04^{\pm 0.84}$ & \cellcolor{highlightorange}$42.36^{\pm 0.21}$ & \cellcolor{highlightorange}$31.96^{\pm 0.04}$ & $1.19$ & $7.33$ \\
\midrule
\multicolumn{7}{c}{\textit{Compression Ratio 0.51}} \\
\midrule
Random & $14.02^{\pm 0.33}$ & $52.08^{\pm 0.49}$ & $39.20^{\pm 0.43}$ & $26.95^{\pm 0.01}$ & $0.97$ & $5.61$ \\
SpeechPrune & $6.52^{\pm 1.14}$  & \cellcolor{highlightorange}$57.39^{\pm 0.74}$ & $41.83^{\pm 0.06}$ & $31.16^{\pm 0.15}$ & $1.14$ & $5.67$ \\
SparseVLM$^{\dagger\ddagger}$ & $9.13^{\pm 1.15}$  & $47.10^{\pm 0.69}$ & $38.80^{\pm 0.27}$ & $31.17^{\pm 0.09}$ & $1.15$ & $6.10$ \\
LTBM & $6.92^{\pm 0.04}$  & $54.58^{\pm 0.42}$ & $40.33^{\pm 0.40}$ & $29.55^{\pm 0.03}$ & $1.15$ & $5.62$ \\
\textbf{AudioTETRIS}$^{\dagger}$ & \cellcolor{highlightorange}$5.77^{\pm 0.33}$ & $57.03^{\pm 0.41}$ & \cellcolor{highlightorange}$42.91^{\pm 0.36}$ & \cellcolor{highlightorange}$31.45^{\pm 0.06}$ & $1.00$ & $6.13$ \\
\midrule
\multicolumn{7}{c}{\textit{Compression Ratio 0.64}} \\
\midrule
Random & $24.08^{\pm 0.26}$ & $50.22^{\pm 1.10}$ & $39.23^{\pm 0.41}$ & $24.64^{\pm 0.02}$ & $0.84$ & $4.47$ \\
SpeechPrune & $8.62^{\pm 0.07}$ & $56.10^{\pm 1.09}$ & $40.30^{\pm 0.69}$ & $30.12^{\pm 0.05}$ & $0.99$ & $4.53$ \\
SparseVLM$^{\dagger\ddagger}$ & $12.40^{\pm 0.55}$ & $47.27^{\pm 1.74}$ & $37.66^{\pm 0.66}$ & $30.79^{\pm 0.12}$ & $0.95$ & $5.08$ \\
LTBM & $8.60^{\pm 0.18}$  & $53.38^{\pm 0.86}$ & $40.70^{\pm 0.78}$ & $29.47^{\pm 0.06}$ & $1.03$ & $4.49$ \\
\textbf{AudioTETRIS}$^{\dagger}$ & \cellcolor{highlightorange}$6.62^{\pm 0.29}$ & \cellcolor{highlightorange}$57.90^{\pm 0.76}$ & \cellcolor{highlightorange}$41.83^{\pm 0.15}$ & \cellcolor{highlightorange}$31.24^{\pm 0.11}$ & $0.86$ & $5.26$ \\
\midrule
\multicolumn{7}{c}{\textit{Compression Ratio 0.75}} \\
\midrule
Random & $31.59^{\pm 0.24}$ & $47.67^{\pm 0.71}$ & $38.57^{\pm 1.16}$ & $22.82^{\pm 0.01}$ & $0.61$ & $3.52$ \\
SpeechPrune & $9.75^{\pm 0.10}$  & $54.17^{\pm 0.92}$ & $40.93^{\pm 0.34}$ & $25.77^{\pm 0.05}$ & $0.69$ & $3.58$ \\
SparseVLM$^{\dagger\ddagger}$ & $17.30^{\pm 0.70}$ & $46.10^{\pm 3.05}$ & $35.83^{\pm 0.12}$ & $30.66^{\pm 0.02}$ & $0.71$ & $4.23$ \\
LTBM & $10.58^{\pm 0.42}$ & $47.84^{\pm 1.05}$ & $39.23^{\pm 0.45}$ & $24.26^{\pm 0.05}$ & $0.70$ & $3.52$ \\
\textbf{AudioTETRIS}$^{\dagger}$ & \cellcolor{highlightorange}$8.50^{\pm 0.18}$ & \cellcolor{highlightorange}$55.34^{\pm 1.25}$ & \cellcolor{highlightorange}$41.46^{\pm 0.32}$ & \cellcolor{highlightorange}$31.33^{\pm 0.07}$ & $0.62$ & $4.32$ \\
\bottomrule
\end{tabular}%
}
\begin{tablenotes}
\footnotesize
\item[$\dagger$] In-LM pruning (token elimination performed inside the LLM backbone).
\item[$\ddagger$] Originally developed for the vision modality, adapted here for speech/audio.
\end{tablenotes}
\end{threeparttable}
\end{table}

\section{Ablation Studies}
\label{sec:ablation}

We isolate the contribution of each component of AudioTETRIS's scoring function and validate the choice of distance metric underlying the transition score \S\ref{sec:method-prior}.

\subsection{Distance Metric for the Transition Score}
\label{sec:ablation-distance-metric}

\begin{figure}[htbp]
\centering
\includegraphics[width=\linewidth]{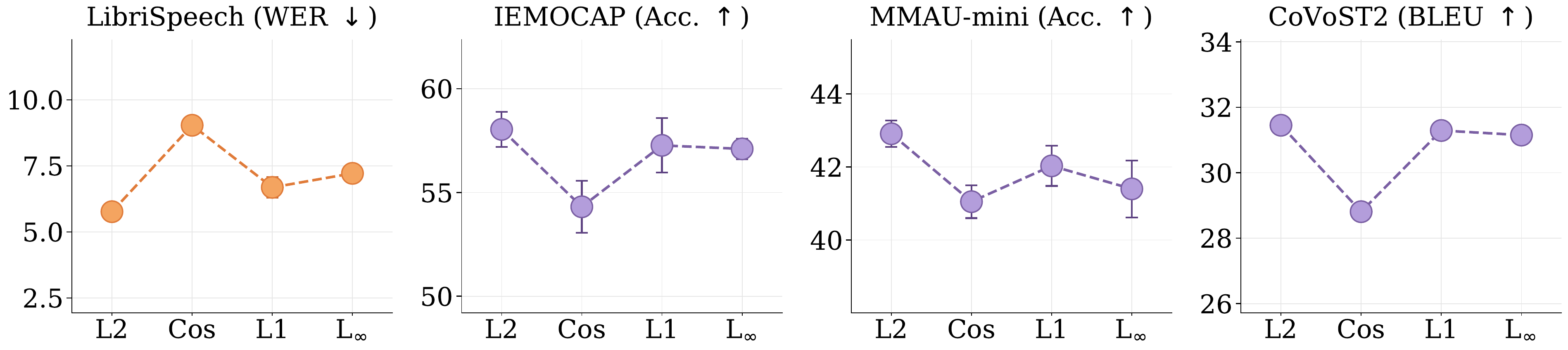}
\caption{WER/accuracy/BLEU across distance metrics for the transition score, at compression ratio $0.51$. Error bars show standard deviation over runs.}
\label{fig:distance_metric_ablation}
\end{figure}

The transition signal $d_t$ requires a distance function to measure the transition magnitude between adjacent encoder hidden states. We compare four standard choices: Euclidean (L2), cosine, Manhattan (L1), and Chebyshev (L$_\infty$) distance, holding all other hyperparameters fixed at a compression ratio of $0.51$. Results are summarized in Figure~\ref{fig:distance_metric_ablation}.

We find that L2 distance is the strongest choice on all four benchmarks, with the largest gap on ASR (cosine increases WER by over 50\%: $9.04\%$ vs.\ $5.77\%$); hence we use it as the default throughout.

\subsection{Component and Stage Ablation Analysis}
\label{sec:ablation-component}

\begin{wrapfigure}{r}{0.5\textwidth}
    \centering
    \vspace{-10pt}
    \includegraphics[width=0.48\textwidth]{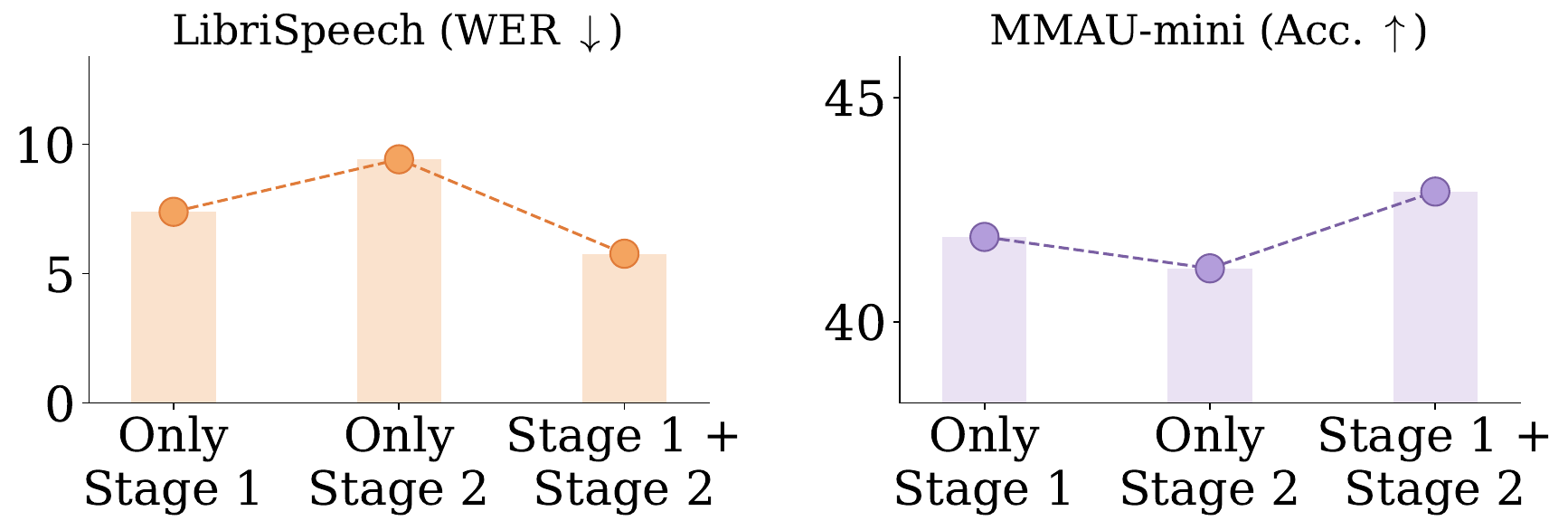}
    \vspace{-8pt}
    \caption{Ablation of AudioTETRIS stages at compression ratio $0.51$.}
    \label{fig:stage_ablation_051}
\end{wrapfigure}

The fused score $s_t^{(\ell)} = \tilde p_t^{(\ell)} \cdot w_t$ (Eq.~\ref{eq:fusion}) combines the depth-adaptive attention term with the acoustic prior. Figure~\ref{fig:stage_ablation_051} isolates each stage's contribution against the full pipeline at compression ratio $0.51$; extended results across all ratios and remaining benchmarks are in Table~\ref{tab:ablation_results} (Appendix~\ref{sec:appendix_ablation}).

We find that both stages combined consistently outperforms either stage alone across all ratios and benchmarks, with the gap widening sharply under aggressive compression (at ratio $0.75$, Stage 2 alone degrades to $15.56\%$ WER vs.\ $8.50\%$ combined). This reflects complementary signals: the phonetic prior $w_t$ preserves phonetically contiguous spans, while the attention term targets task relevance - most visible on LibriSpeech, where ASR's sensitivity to fine-grained transitions makes the two-stage advantage largest.

\section{Conclusion}
\label{sec:conclusion}

We presented AudioTETRIS, a training-free framework for compressing audio tokens in LALMs. By leveraging the audio encoder's hidden-state dynamics as a phoneme-transition prior and fusing it with depth-adaptive decoder attention, AudioTETRIS preserves acoustically rich and semantically critical regions without requiring auxiliary training or fine-tuning. Experiments across Qwen2-Audio, Phi-4-Multimodal, and Audio Flamingo 3, demonstrate that AudioTETRIS retains near-lossless performance at light compression, while widening its advantage over training-free baselines as compression gets aggressive. Furthermore, our findings highlight that ASR is highly sensitive to token loss making speech-aware structural priors essential for exact lexical recovery. Ultimately, this work highlights a broader principle for multimodal architecture design and token-compression methods: intermediate representations computed by modality-specific encoders contain valuable structural geometry that should be propagated forward rather than discarded at the LLM interface.

\subsubsection*{Limitations}
AudioTETRIS's acoustic prior assumes speech-encoder hidden states exhibit large-norm transitions at phonetic boundaries - validated for Whisper-based encoders (Qwen2-Audio, Audio Flamingo 3) and the Conformer-based Phi-4-Multimodal encoder, but not confirmed for encoders trained with different objectives or aggressive temporal downsampling (e.g., HTSAT-style audio-event encoders). Second, while the early pruning layer is fixed by construction, the later crossover layer $\ell^\dagger$ requires a short backbone-specific calibration pass (Appendix~\ref{sec:appendix-crossover}) - a one-time cost, but the method is not fully zero-shot to unseen architectures without it. Third, because pruning happens progressively inside the decoder rather than once before prefill, AudioTETRIS carries a small TFLOP overhead relative to single-shot methods (\S\ref{sec:exp-efficiency}), though this is largely offset by its accuracy gains. Finally, our evaluation spans four English-source benchmarks; results on noisier, or multilingual audio remain untested, and pushing compression beyond the rates studied here would likely erode lexical accuracy first, given ASR's sensitivity to aggressive compression.

\newpage
\subsection*{Acknowledgements}
Funded by the European Union. Views and opinions expressed are however those of the author(s) only and do not necessarily reflect those of the European Union or the European Commission. Neither the European Union nor the European Commission can be held responsible for them. This work was supported by the European Union’s Horizon Europe research and innovation programme under grant agreement No 101214398 (ELLIOT). This publication is part of project 2026.008 of the Computing Time National Computing Facilities program which is financed by the Dutch Research Council (NWO) under the grant https://doi.org/10.61686/WBFDE48237.

\subsection*{Reproducibility Statement}
AudioTETRIS is training-free, so no training procedure, optimizer, or 
learning-rate schedule needs to be specified for reproduction. All 
components needed to reproduce our results are given in the main text 
and appendix: the phoneme transition score (Eq.~\ref{eq:dt}-\ref{eq:variance-weight}), the depth-adaptive attention score and its automated crossover-layer discovery procedure (Eq.~\ref{eq:ptilde}-\ref{eq:crossover-layer}, Appendix~\ref{sec:appendix-crossover}). A one-time, per-backbone calibration pass (Appendix~\ref{sec:appendix-crossover}) locates the crossover layer $\ell^\dagger$ from attention statistics; this involves no gradient computation or parameter updates. The full fused-score pruning schedule are summarized as pseudocode in Algorithms~\ref{alg:calibration} and~\ref{alg:tetris} (Appendix~\ref{sec:appendix-algorithm}). Pruning layers, per-stage retention rates, and the closed-form multi-stage schedule are specified in \S\ref{sec:exp-setup} and Appendix~\ref{app:multi-stage} (Eq.~\ref{eq:multistage-closed-form}, Table~\ref{tab:multistage-combo-specifics}). Backbone checkpoints, exact dataset splits, and their licenses are listed in Appendix~\ref{sec:appendix-licenses} (Tables~\ref{tab:model-licenses}, \ref{tab:dataset-licenses}); alignment sources used for the boundary-correlation analysis (\S\ref{sec:analysis-prior}) are described in Appendix~\ref{sec:appendix-covost-iemocap}.  We will release our full implementation, including the pruning modules described in Algorithms~\ref{alg:calibration} and~\ref{alg:tetris} and the evaluation scripts used to produce Tables~\ref{tab:main_results}-\ref{tab:audioflamingo3_appendix}, at \url{https://github.com/ShreyaSajid/AudioTETRIS}.

\bibliography{iclr2027_conference}
\bibliographystyle{iclr2027_conference}

\newpage
\appendix
\section{Appendix}

\subsection{Implementation Details}
\subsubsection{Backbones and Checkpoints}
\label{sec:appendix-licenses}

Table~\ref{tab:model-licenses} lists the exact checkpoints and licenses for the three backbones evaluated in this work. All experiments use unmodified pretrained weights; the only backbone-side code changes are load-time monkey-patches to the decoder forward pass implementing AudioTETRIS's pruning (\S\ref{sec:method}). No checkpoint weights were altered or fine-tuned.

\begin{table}[h]
\centering
\caption{Pretrained checkpoints used in this work.}
\label{tab:model-licenses}
\begin{threeparttable}
\resizebox{0.95\textwidth}{!}{%
\begin{tabular}{@{}lll@{}}
\toprule
\textbf{Backbone} & \textbf{Checkpoint} & \textbf{License} \\
\midrule
Qwen2-Audio \citep{qwen2audio} & \texttt{Qwen/Qwen2-Audio-7B-Instruct} & Apache 2.0 \\
Phi-4-Multimodal \citep{microsoft2025phi4minitechnicalreportcompact} & \texttt{microsoft/Phi-4-multimodal-instruct} & MIT \\
Audio Flamingo 3 \citep{goel2025audioflamingo3advancing} & \texttt{nvidia/audio-flamingo-3-hf} & NVIDIA OneWay Noncommercial License\tnote{*} \\
\bottomrule
\end{tabular}%
}
\begin{tablenotes}
\footnotesize
\item[*] Non-commercial research use only; see checkpoint card for full terms.
\end{tablenotes}
\end{threeparttable}
\end{table}
\vspace{-0.5em}

\subsubsection{Datasets, Splits, and Evaluation}

Table~\ref{tab:dataset-licenses} lists the datasets used for evaluation, the exact splits, and their licensing. We accessed all four via the \texttt{AudioLLMs} organization's re-packaged mirrors on the Hugging Face Hub, used throughout the AudioBench evaluation harness~\citep{AudioBench2025}, rather than the original distribution channels for each corpus. For the CoVoST2 phoneme alignments generated for Appendix~\ref{sec:appendix-covost-iemocap}, these are derivative annotations produced by us using the Montreal Forced Aligner and are not redistributed third-party material.

\begin{table*}[h]
\centering
\caption{Evaluation datasets, splits, and licensing.}
\label{tab:dataset-licenses}
\resizebox{0.95\textwidth}{!}{%
\begin{tabular}{@{}lll@{}}
\toprule
\textbf{Dataset} & \textbf{Split used} & \textbf{Original license} \\
\midrule
LibriSpeech (test-other) \citep{panayotov2015librispeech} & \texttt{AudioLLMs/librispeech\_test\_other}, \texttt{test} & CC BY 4.0 \\
IEMOCAP \citep{iemocap} & \texttt{AudioLLMs/iemocap\_emotion\_recognition}, \texttt{test} & Custom non-commercial research license (USC SAIL) \\
MMAU-mini \citep{MMAU2024} & \texttt{AudioLLMs/MMAU-mini}, \texttt{test} & Not explicitly stated by source \\
CoVoST2 (En$\to$Zh) \citep{wang2020covost2massivelymultilingual} & \texttt{AudioLLMs/covost2\_en\_zh\_test}, \texttt{test} & CC0 (per original release); Hub mirror tagged CC BY-NC 4.0 \\
\bottomrule
\end{tabular}%
}
\end{table*}

\subsubsection{SparseVLM Adaptation to Fixed-Rate Speech Pruning}
\label{sec:appendix-sparsevlm}
SparseVLM~\citep{zhang2025sparsevlmvisualtokensparsification} originally proposed an adaptive, dynamic pruning schedule based on the rank of the visual-text attention matrix. However, its public release\footnote{\url{https://github.com/Gumpest/SparseVLMs}, accessed \texttt{May 2026}, commit \texttt{87fe4319430e079a778c25d4ddf4588a3c4038ca 2025-12-22}.} replaces this dynamic controller with static token-retention schedules. Because the dynamic rank-based ratio is absent from all usable releases (\texttt{main}, \texttt{v1.5}, \texttt{v2.0}), we could not evaluate SparseVLM’s originally described adaptive deployment directly. Instead, we adapt its public attention-guided scoring mechanism to our fixed-rate setup for a direct, controlled comparison. 

\textbf{Adaptation.} We retain SparseVLM’s core scoring mechanism - using the text raters, and text-token attention to rate audio token importance - and apply it at layers $\{1, 15\}$ (Appendix~\ref{sec:appendix-early-layer}). At each stage, we keep the top fraction $r$ of audio tokens by score to match our evaluated compression targets ($0.36, 0.51, 0.64, 0.75$). This adaptation isolates the impact of SparseVLM’s attention-based importance metric under identical layer placements and compute budgets. 

\subsubsection{Attention Backend for Score-Based Pruning}

\label{app:attn-backend}

AudioTETRIS and the attention-score-based baselines we compare against
(SparseVLM~\cite{zhang2025sparsevlmvisualtokensparsification}, SpeechPrune \citep{speechprune}) all derive token-importance scores from attention probabilities between query and audio tokens. Fused attention kernels such as FlashAttention-2~\citep{dao2024flashattention2} and PyTorch's scaled-dot-product attention (SDPA) do not materialize these probabilities: both return \texttt{attn\_weights=None}, since the attention matrix is never explicitly formed. Eager attention is the only implementation that exposes the full $[B, H, N^{(\ell)}, N^{(\ell)}]$ probability tensor required for scoring.

We therefore run the attention backend in eager mode only at the
layers where token scores are computed, and restore the model's
configured kernel (FlashAttention-2 \citep{dao2024flashattention2}, in our setup) for the forward
pass that produces the hidden states and KV cache used downstream.
This substitution is local to the small set of scoring layers
(\S\ref{sec:method}) rather than applied model-wide, so the
FlashAttention-2 fast path is preserved for the remainder of the
network's depth. We apply this identically across AudioTETRIS and all
attention-score-based baselines, so the latency and FLOPs figures
reported in \S\ref{sec:exp-main} reflect the same hybrid execution
path for every method and remain directly comparable.

We note that this eager-attention fallback is a memory-bandwidth-bound cost that is not reflected in the TFLOPs column, which counts compute operations only; it is, however, fully captured in our reported wall-clock latency, since all latency numbers are measured end-to-end on the actual hybrid execution path (eager at scoring layers, FlashAttention-2 elsewhere). Because the eager fallback is confined to exactly two decoder layers per backbone (layer 1 and that backbone's calibrated $\ell^{\dagger}$; \S\ref{sec:exp-setup}), its contribution to total latency is small in practice, and it could be eliminated entirely in a production deployment with a custom fused kernel that computes only the row/column marginal attention sums our scoring requires, without materializing the full attention matrix - a one-time kernel-engineering cost rather than a potential structural limitation of the method.

\subsection{Method Details and Analysis}
\subsubsection{Algorithmic Summary}

\label{sec:appendix-algorithm}

For reproducibility, Algorithms~\ref{alg:calibration} and~\ref{alg:tetris} summarize the full pipeline described in \S\ref{sec:method}. Algorithm~\ref{alg:calibration} is a one-time, offline calibration step run once per backbone to locate the crossover layer $\ell^{\dagger}$ (\S\ref{sec:appendix-crossover}); Algorithm~\ref{alg:tetris} is the training-free inference procedure applied per input, using $\ell^{\dagger}$ from calibration to determine which attention branch of Eq.~\ref{eq:ptilde} to use at each pruning layer.

\begin{algorithm}[h]
\caption{Crossover Layer Calibration (offline, run once per backbone)}
\label{alg:calibration}
\begin{algorithmic}[1]
\Require Calibration set $\mathcal{U}$ (audio-query pairs), backbone with $L_{\text{dec}}$ decoder layers
\Ensure Crossover layer $\ell^{\dagger}$
\For{each utterance $u \in \mathcal{U}$}
    \State Run forward pass; record attention tensors $A^{(\ell)}_u$ for all layers $\ell = 1, \dots, L_{\text{dec}}$
\EndFor
\For{$\ell = 1, \dots, L_{\text{dec}}$}
    \State $a_{\text{intra}}(\ell) \gets$ mean attention mass from $\mathcal{Q}_a$ rows onto audio tokens (Eq.~\ref{eq:ptilde}, upper case), averaged over $\mathcal{U}$
    \State $a_{\text{cross}}(\ell) \gets$ mean attention mass from $\mathcal{Q}_q$ rows onto audio tokens (Eq.~\ref{eq:ptilde}, lower case), averaged over $\mathcal{U}$
\EndFor
\State $\ell^{\dagger} \gets \min\{\ell : a_{\text{cross}}(\ell) \ge a_{\text{intra}}(\ell)\}$ \Comment{Eq.~\ref{eq:crossover-layer}}
\State \Return $\ell^{\dagger}$
\end{algorithmic}
\end{algorithm}

\begin{algorithm}[h]
\caption{AudioTETRIS Inference (training-free, applied per input)}
\label{alg:tetris}
\begin{algorithmic}[1]
\Require Audio encoder hidden states $H = \{h_1, \dots, h_T\}$, pruning layers $\{\ell_1, \ell_2\}$ with $\ell_1 < \ell^{\dagger} \le \ell_2$ (in practice $\ell_2 = \ell^{\dagger}$, the backbone's calibrated crossover layer; \S\ref{sec:exp-setup}), target retention fractions $\{k_{\ell_1}, k_{\ell_2}\}$, decoder with $L_{\text{dec}}$ layers
\Ensure Compressed decoder output

\Statex \textit{\textbf{Stage 1: Acoustic Transition Prior} (computed once, before decoding)}
\State $d_1 \gets 0$
\For{$t = 2, \dots, T$}
    \State $d_t \gets \|h_t - h_{t-1}\|_2^2$ \Comment{Eq.~\ref{eq:dt}}
\EndFor
\State $\bar{d} \gets \frac{1}{T}\sum_{i=1}^{T} d_i$
\For{$t = 1, \dots, T$}
    \State $w_t \gets d_t / \bar{d}$ \Comment{Eq.~\ref{eq:variance-weight}}
\EndFor

\Statex \textit{\textbf{Stage 2: Depth-Adaptive Pruning} (during decoder forward pass)}
\State $\mathcal{T} \gets \{1, \dots, T\}$ \Comment{surviving audio-token indices}
\For{$\ell = 1, \dots, L_{\text{dec}}$}
    \State Run decoder layer $\ell$ on current sequence (audio tokens indexed by $\mathcal{T}$, plus query tokens)
    \If{$\ell \in \{\ell_1, \ell_2\}$}
        \If{$\ell < \ell^{\dagger}$}
            \State $\tilde{p}_t^{(\ell)} \gets$ mean attention from audio-query rows $\mathcal{Q}_a$ onto token $t$, $\forall t \in \mathcal{T}$ \Comment{Eq.~\ref{eq:ptilde}, intra-audio}
        \Else
            \State $\tilde{p}_t^{(\ell)} \gets$ mean attention from text-query rows $\mathcal{Q}_q$ onto token $t$, $\forall t \in \mathcal{T}$ \Comment{Eq.~\ref{eq:ptilde}, cross-modal}
        \EndIf
        \State $s_t^{(\ell)} \gets \tilde{p}_t^{(\ell)} \cdot w_t, \quad \forall t \in \mathcal{T}$ \Comment{Eq.~\ref{eq:fusion}}
        \State $\mathcal{T} \gets$ indices of the top $\lceil k_\ell \cdot |\mathcal{T}| \rceil$ tokens in $\mathcal{T}$ ranked by $s_t^{(\ell)}$
        \State Evict pruned tokens from hidden states and KV cache; update RoPE position indices for $\mathcal{T}$
    \EndIf
\EndFor
\State \Return final decoder output
\end{algorithmic}
\end{algorithm}

\subsubsection{Pruning Layer Selection}

\label{app:layer-selection}

AudioTETRIS prunes at two decoder layers (\S\ref{sec:exp-setup}), and the two are chosen by different criteria because they serve different roles. The early layer is about saving compute: it is fixed by design as the earliest depth at which the attention term exists, and pruning there is viable because the acoustic prior compensates for weak early attention. The late layer $\ell^\dagger$ is about task relevance: it is calibrated per backbone from attention crossover statistics to locate where cross-modal grounding takes over (\S\ref{sec:appendix-late-layer}). We motivate each choice separately below.

\paragraph{Early Pruning Layer.}
\label{sec:appendix-early-layer}
 
Because self-attention cost accrues at every subsequent decoder layer, the earlier a token is removed, the larger the cumulative FLOP reduction across the rest of the forward pass. We therefore place the first pruning stage at layer 1 - the earliest layer at which the depth-adaptive attention score $\tilde p_t^{(\ell)}$ (Eq.~\ref{eq:ptilde}) is defined, since it requires at least one decoder self-attention computation to exist. Deferring pruning to a deeper early layer might yield a marginally sharper intra-audio attention signal, but this benefit is outweighed by the compute lost across every additional layer the full, unpruned token sequence has to pass through.
 
Early pruning works well for AudioTETRIS because the layer-1 score does not rely on attention alone. The fused score $s_t^{(1)} = \tilde{p}_t^{(1)} \cdot w_t$ (Eq.~\ref{eq:fusion}) combines the attention-based importance $\tilde{p}_t^{(1)}$, computed from a single decoder layer, with the acoustic transition prior $w_t$ (Eq.~\ref{eq:variance-weight}). The prior comes directly from the audio encoder's hidden states, so it is already fully informative before the decoder even starts (\S\ref{sec:analysis-prior}).

The prior's independence from decoder depth matters because attention alone is a weak signal at layer 1, where the decoder has not yet built up much context. The acoustic prior compensates for this weakness, making the fused score far more reliable at this depth than attention alone would be. This reliability is what lets us prune aggressively at layer 1 without the accuracy collapse that a purely attention-based method suffers. Our component ablation supports this conclusion (\S\ref{sec:ablation-component}, Table~\ref{tab:ablation_results}): `Only Stage 2' (attention alone) degrades much faster as compression increases than either `Only Stage 1' or the full fused score.

\paragraph{Late Pruning Layer: Crossover Calibration.}
\label{sec:appendix-crossover}
\label{sec:appendix-late-layer}
 
Section~\ref{sec:method-attn} defines the crossover layer $\ell^{\dagger}$ (Eq.~\ref{eq:crossover-layer}) as the first decoder depth at which cross-modal attention mass overtakes intra-audio attention mass, and uses it to set the layer partition $\{\mathcal{L}_{\text{early}}, \mathcal{L}_{\text{later}}\}$ for the depth-adaptive attention score $\tilde{p}_t^{(\ell)}$ (Eq.~\ref{eq:ptilde}). Here we describe the calibration procedure used to estimate $\ell^{\dagger}$ and report the resulting per-layer attention statistics for each backbone.
 
\textbf{Calibration set.} We estimate $\ell^{\dagger}$ from 150 utterances drawn from LibriSpeech test-other~\citep{panayotov2015librispeech}, disjoint from the utterances used for downstream evaluation in \S\ref{sec:exp-main}. For each calibration utterance, we run a forward pass through the target backbone and record the full decoder attention tensors $A^{(\ell)}$ at every layer, with no gradient updates or parameter changes. Calibration takes under 4 minutes on a single A100 and is amortized across all subsequent inputs for that backbone.

\textbf{Per-layer attention mass.} For each decoder layer $\ell$, we define the mean intra-audio and cross-modal attention mass, restating the two branches of Eq.~\ref{eq:ptilde} without the depth partition:
\begin{align}
a_{\text{intra}}(\ell) &= \frac{1}{|\mathcal{U}|}\sum_{u \in \mathcal{U}} \frac{1}{H \cdot |\mathcal{Q}_a| \cdot T_u} \sum_{h=1}^{H} \sum_{i \in \mathcal{Q}_a} \sum_{t=1}^{T_u} A^{(\ell)}_{h,i,t}, \\
a_{\text{cross}}(\ell) &= \frac{1}{|\mathcal{U}|}\sum_{u \in \mathcal{U}} \frac{1}{H \cdot |\mathcal{Q}_q| \cdot T_u} \sum_{h=1}^{H} \sum_{i \in \mathcal{Q}_q} \sum_{t=1}^{T_u} A^{(\ell)}_{h,i,t},
\end{align}
where $\mathcal{U}$ is the calibration set, $T_u$ is the number of audio tokens in utterance $u$, and $\mathcal{Q}_a$, $\mathcal{Q}_q$ are the audio-query and text-query row index sets as in Eq.~\ref{eq:ptilde}. The crossover layer is then $\ell^{\dagger} = \min\{\ell : a_{\text{cross}}(\ell) \ge a_{\text{intra}}(\ell)\}$, matching Eq.~\ref{eq:crossover-layer}.
 
\begin{figure}[h]
    \centering
    \includegraphics[width=\textwidth]{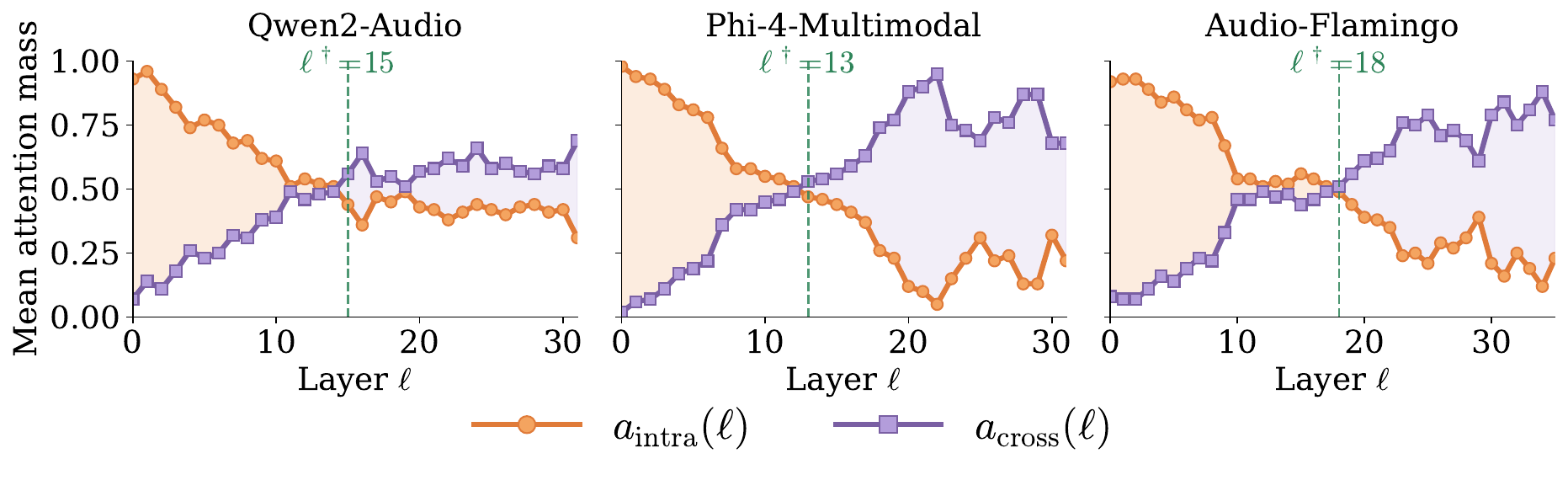}
    \caption{Per-layer intra-audio ($a_{\text{intra}}(\ell)$) and cross-modal ($a_{\text{cross}}(\ell)$) attention mass on the LibriSpeech test-other calibration subset, for Qwen2-Audio \citep{qwen2audio}, Phi-4-Multimodal \citep{microsoft2025phi4minitechnicalreportcompact}, and Audio Flamingo 3 \citep{goel2025audioflamingo3advancing} The dashed vertical line marks the identified crossover layer $\ell^{\dagger}$ for each backbone. Intra-audio attention dominates in early layers and is progressively overtaken by cross-modal attention as depth increases, consistent with Figure~\ref{fig:attn_split_pies}.}
    \label{fig:crossover_calibration}
\end{figure}
 
\textbf{Results.} As shown in Figure~\ref{fig:crossover_calibration}, all three backbones exhibit the same qualitative pattern: intra-audio attention mass is highest in early layers and declines monotonically (on average) with depth, while cross-modal attention mass rises correspondingly, crossing over in the middle third of the network. The calibration procedure identifies $\ell^{\dagger} = 15$ for Qwen2-Audio, consistent with the pruning schedule used in \S\ref{sec:exp-setup}, $\ell^{\dagger} = 13$ for Phi-4-Multimodal, and $\ell^{\dagger} = 18$ for Audio Flamingo 3. The crossover layer therefore shifts across backbones with differing encoder-decoder depth and subsampling rates.

\subsubsection{Transition Prior Computation Pipeline}

\label{sec:appendix-boundary-schematic}

\begin{figure}[h]
    \centering
    \includegraphics[width=\textwidth]{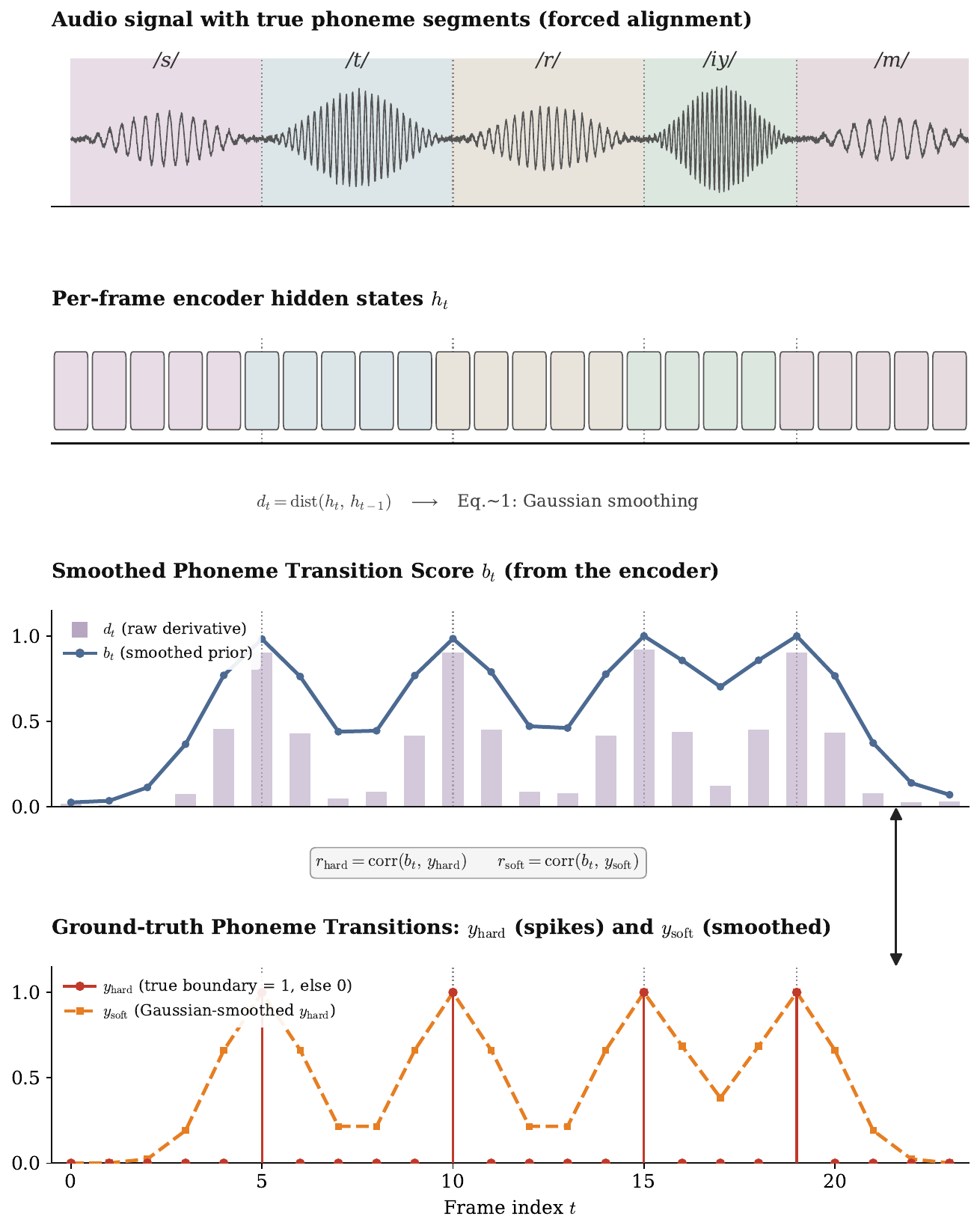}
    \caption{Schematic of the phoneme-transition-aware prior pipeline. \textbf{Top:} audio waveform with forced-alignment phoneme segments. \textbf{Second row:} per-token encoder hidden states $h_t$, colored by phoneme membership. \textbf{Third row:} raw derivative $d_t$ (bars) and the Gaussian-smoothed transition score $b_t$ (line). \textbf{Bottom:} ground-truth transition labels, shown as discrete spikes $y_{\text{hard}}$ and their Gaussian-smoothed form $y_{\text{soft}}$, used to compute $r_{\text{hard}}$ and $r_{\text{soft}}$.}
    \label{fig:boundary-schematic}
\end{figure}

Figure~\ref{fig:boundary-schematic} illustrates the end-to-end computation of the phoneme transition score $b_t$ and its relationship to ground-truth phoneme transitions, using a toy example spanning the phonemes \textit{/s/-/t/-/r/-/iy/-/m/}. The raw derivative $d_t$ (Eq.~\ref{eq:dt}) is computed from consecutive encoder hidden states, producing sparse spikes near true phonetic transitions but with substantial token-to-token noise. Gaussian smoothing (Eq.~\ref{eq:bt}) converts this into the continuous transition prior $b_t$, which we correlate against both the discrete ground-truth transition indicator $y_{\text{hard}}$ and its Gaussian-smoothed counterpart $y_{\text{soft}}$ to obtain $r_{\text{hard}}$ and $r_{\text{soft}}$ respectively.

\subsubsection{Boundary Correlation on CoVoST2 and IEMOCAP}

\label{sec:appendix-covost-iemocap}

Section~\ref{sec:analysis-prior} reports boundary-correlation results on LibriSpeech (Figure~\ref{fig:layer_correlations}), using the MFA-based
forced alignments made available through \citep{librispeech_alignments_github}. Here we repeat the same analysis on our two other evaluation datasets, CoVoST2 (speech translation) \citep{wang2020covost2massivelymultilingual}
and IEMOCAP (emotion recognition) \citep{iemocap}, to confirm that the mid-network
correlation peak is not an artifact of LibriSpeech's read-speech domain.

\textbf{Alignment sources.} IEMOCAP ships with official word-, syllable-, and phoneme-level forced alignments as part of its dataset release, generated with Sphinx-III~\citep{iemocap}; we use these directly. No comparable off-the-shelf alignment resource exists for CoVoST2, so we
generated phoneme alignments ourselves using the Montreal Forced Aligner (MFA) \citep{mcauliffe2017montreal} on CoVoST2's English-source audio and transcripts, matching the
alignment methodology used for the LibriSpeech release to keep the comparison consistent across datasets.\footnote{IEMOCAP's original release
notes that alignment accuracy was screened but not formally evaluated by the corpus authors; we flag this as a caveat on alignment quality relative to the MFA-based LibriSpeech and CoVoST2 alignments used elsewhere in this work.} For both datasets we recompute $b_t$ per layer and correlate against the same $y_{\text{hard}}$/$y_{\text{soft}}$ construction as in Section~\ref{sec:analysis-prior}, holding $\sigma=2$ fixed.

\begin{figure}[t]
    \centering
    \includegraphics[width=\textwidth]{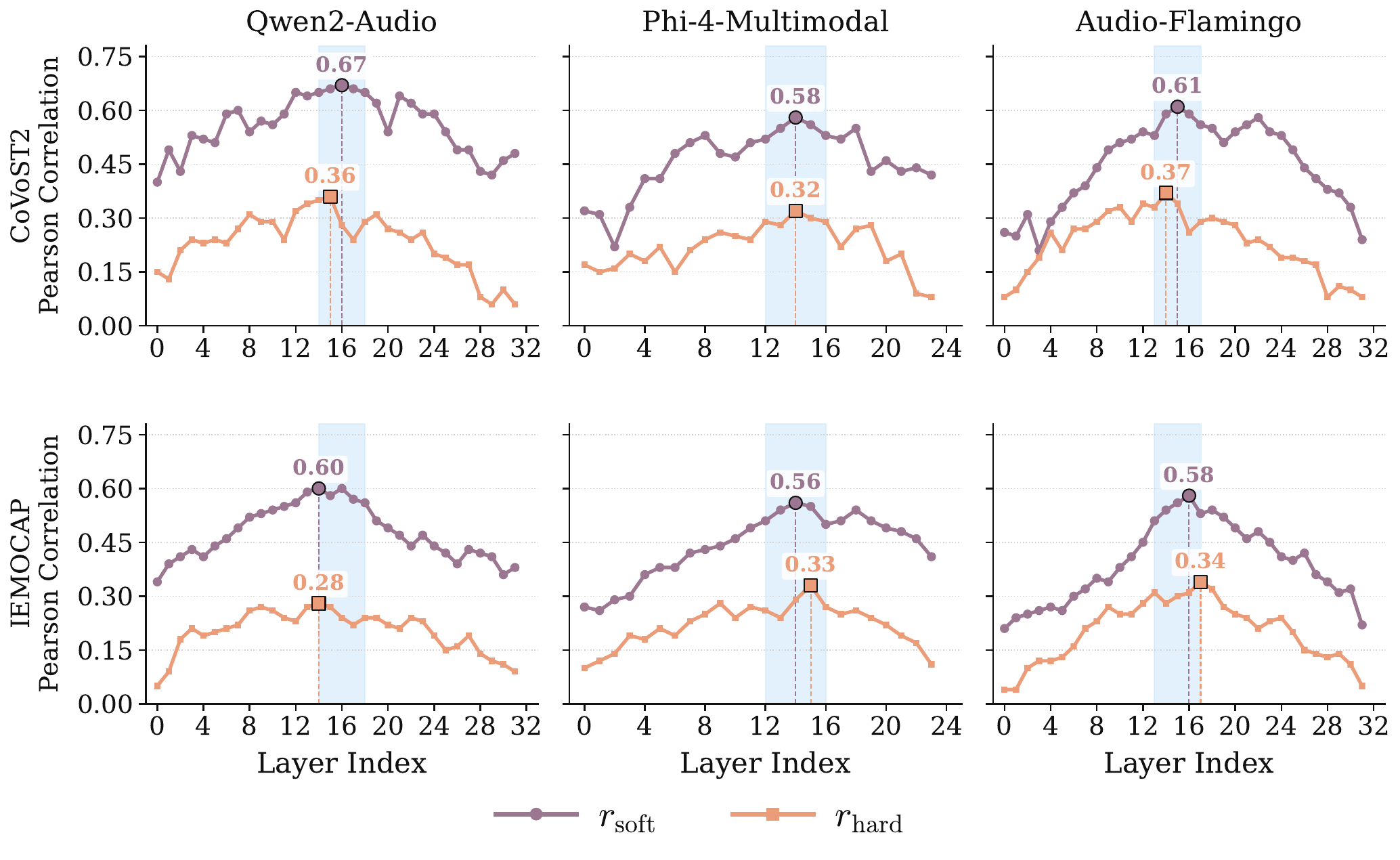}
    \caption{
        Layer-wise boundary correlation ($r_{\text{soft}}$, $r_{\text{hard}}$)
        across Qwen2-Audio, Phi-4-Multimodal, and Audio Flamingo 3, computed on
        CoVoST2 (\textbf{top}, MFA alignments generated in-house) and IEMOCAP
        (\textbf{bottom}, official released alignments). Shaded regions mark
        each backbone's mid-network sweet spot, consistent with
        Figure~\ref{fig:layer_correlations}. Dashed vertical lines and
        annotated values mark the peak layer for each curve.
    }
    \label{fig:covost_iemocap_correlations}
\end{figure}

As shown in Figure~\ref{fig:covost_iemocap_correlations}, both datasets reproduce the mid-network peak observed on LibriSpeech across all three backbones, with $r_{\text{soft}}$ peaking in the same layer range identified
by the sweet-spot shading (shared with Figure~\ref{fig:layer_correlations}).
Peak $r_{\text{soft}}$ is consistently higher on CoVoST2 than on IEMOCAP for every backbone (e.g., Qwen2-Audio: $0.67$ vs.\ $0.60$; Audio Flamingo 3: $0.61$ vs.\ $0.58$), which is consistent with our finding in Section~\ref{sec:exp-main} that fine-grained lexical tasks depend more heavily on precise boundary localization than paralinguistic classification does.

\subsubsection{Sensitivity to the Smoothing Width $\sigma$}

\label{sec:appendix_sigma_ablation}

Eq.~\ref{eq:bt} defines the transition prior $b_t$ by convolving the raw distance
signal $d_t$ with a Gaussian kernel of width $\sigma$, and we apply the same
$\sigma$ to smooth the discrete forced-alignment labels $y_{\text{hard}}$ into
the continuous target $y_{\text{soft}}$. In the main text we fix $\sigma=2$
(encoder tokens). Here we ablate this choice to characterize how $r_{\text{soft}}$ and $r_{\text{hard}}$ vary as $\sigma$ moves away from the selected smoothing width.

\begin{figure}[t]
    \centering
    \includegraphics[width=0.9\textwidth]{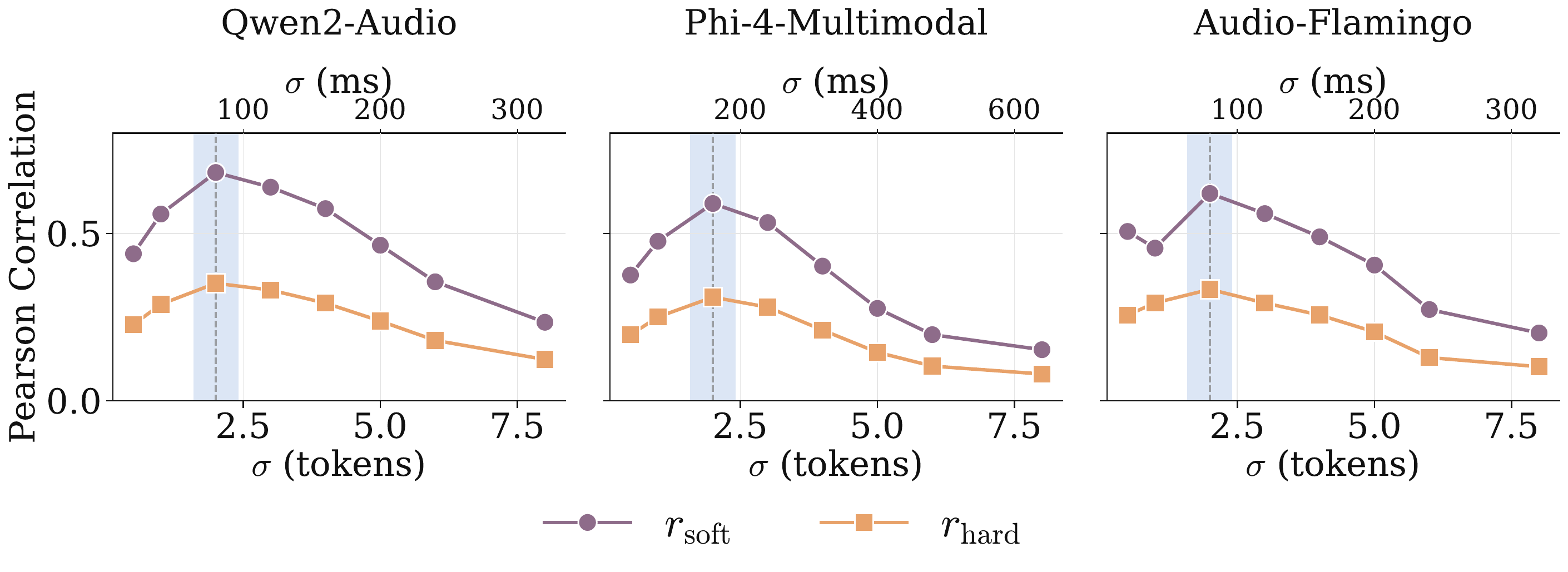}
    \caption{
        Sensitivity of the transition-prior correlation to the Gaussian
        smoothing width $\sigma$, on all backbones. Each panel sweeps $\sigma$ (encoder tokens, with milliseconds on the secondary top axis) and reports $r_{\text{soft}}$ and $r_{\text{hard}}$ at that backbone's mid-network sweet-spot layer. The shaded band and dashed line mark $\sigma=2$, the value used throughout the main text. 
    }
    \label{fig:sigma_ablation}
\end{figure}

We sweep $\sigma \in \{0.5, 1, 2, 3, 4, 5, 6, 8\}$
encoder tokens and recompute $r_{\text{soft}}$ and $r_{\text{hard}}$ at each
value, holding the layer fixed at the mid-network sweet spot identified in
Section~\ref{sec:analysis-prior} for each backbone. Because Phi-4-Multimodal's
audio encoder subsamples more aggressively than Qwen2-Audio's or
Audio Flamingo's, the same $\sigma$ in token
units corresponds to roughly twice the real-time smoothing window for
Phi-4-Multimodal; Figure~\ref{fig:sigma_ablation} therefore also reports
$\sigma$ in milliseconds on a secondary axis for each backbone.

As shown in Figure~\ref{fig:sigma_ablation}, both $r_{\text{soft}}$ and $r_{\text{hard}}$ consistently exhibit an inverted U-shaped curve across all backbones, peaking cleanly at $\sigma = 2$ tokens. Performance degrades on either side: smaller values fail to account for context distribution, while larger values ($\sigma > 2$) excessively blur distinct phonetic transitions.

\subsubsection{Qualitative Encoder Embedding Trajectories}

\label{section:appendix-trajectory}
\begin{figure}[t]
    \centering
    \vspace{-0.5em}
    \includegraphics[width=0.85\textwidth]{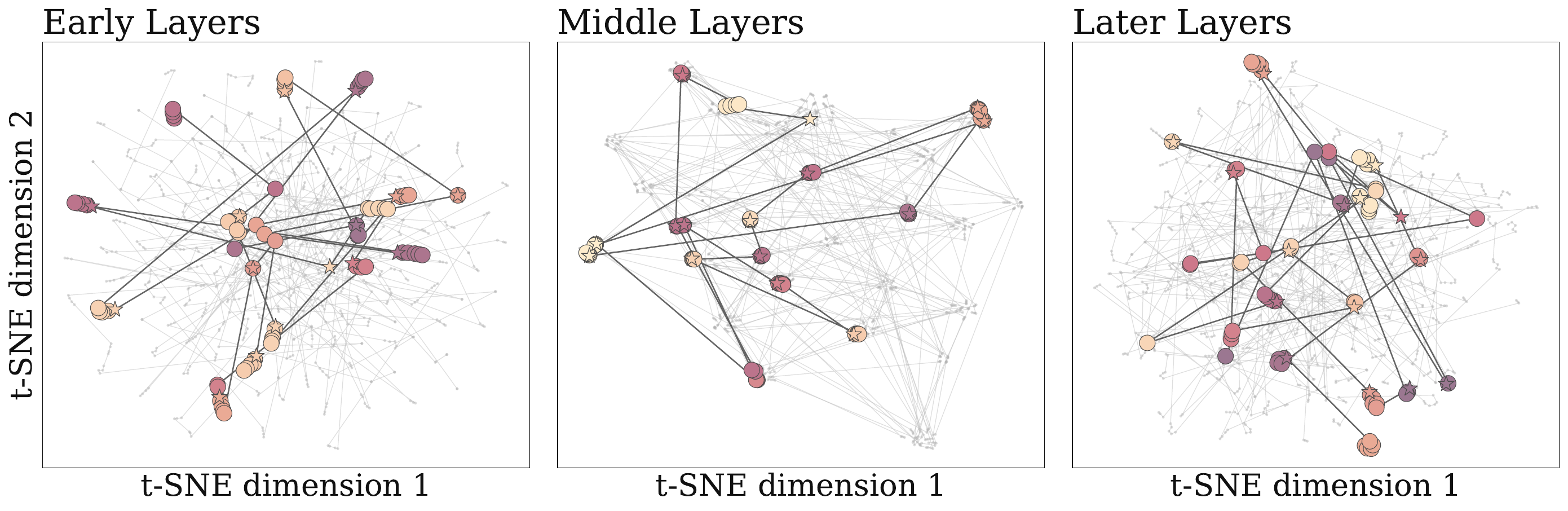}
    \caption{
        Qualitative analysis of Qwen2-Audio encoder token embedding trajectories for a single LibriSpeech utterance, projected via t-SNE and colored by phoneme identity, at early, middle, and late encoder depths. Colored points \phonemedot{} denote individual tokens where different colors represent different phonemes; Star markers \starmark{} indicate frames at phoneme transitions per forced alignment; Faint gray points \graydot{} show other utterances' frames for spatial context. 
    }
    \label{fig:embedding_trajectory}
\end{figure}

Figure \ref{fig:embedding_trajectory} complements the quantitative diagnostics of Section~\ref{sec:analysis-prior} with a qualitative view of the same phenomenon. We project the per-token encoder hidden states $h_t$ of a single LibriSpeech test-other utterance into two dimensions with t-SNE at an early (layer 4), middle (layer 16), and late encoder layer (layer 28) of Qwen2-Audio, and trace the utterance's token sequence through the projected space. Frames from other utterances are shown in grey to provide a shared spatial reference.

At the middle layer, which coincides with the peak of both $r_{\text{soft}}$ and kurtosis (Figure~\ref{fig:layer_correlations}), tokens belonging to the same phoneme form compact clusters. At the early and late layers, tokens are less consolidated by phoneme identity and boundary frames are less clearly separated from within-segment frames, consistent with the lower correlation and kurtosis observed at those depths. Since t-SNE preserves local neighbourhoods but not global distances, we treat this figure as illustrative only. The quantitative evidence for the mid-network sweet spot is given by the correlation and kurtosis analyses in Section~\ref{sec:analysis-prior} and Appendix~\ref{sec:appendix-covost-iemocap}.

\subsection{Additional Experiments}
\subsubsection{Extended Results on Phi-4-Multimodal and Audio Flamingo 3}

\label{sec:appendix_multibackbone}

Section~\ref{sec:exp-main} and Figure~\ref{fig:wer_mmau_lollipop} reports results on Phi-4-Multimodal and Audio Flamingo 3 at a single compression ratio ($0.51$) for brevity. Here we provide the extended results for both backbones across all four evaluated compression ratios ($0.36$, $0.51$, $0.64$, $0.75$), reported in Table~\ref{tab:phi4_appendix} for Phi-4-Multimodal and Table~\ref{tab:audioflamingo3_appendix} for Audio Flamingo 3. Consistent with the Qwen2-Audio results in Table~\ref{tab:main_results}, AudioTETRIS maintains near-lossless performance across both additional backbones at low compression and shows a widening advantage over baselines as compression increases, indicating that the method generalizes across encoder-decoder architectures rather than being tuned to a single backbone.

\begin{table}[t]
\centering
\caption{Extended results for Phi-4-Multimodal across all compression ratios.}
\label{tab:phi4_appendix}
\begin{threeparttable}
\resizebox{\textwidth}{!}{%
\begin{tabular}{@{}l c c c c c c@{}}
\toprule
\textbf{Method} & \textbf{LibriSpeech WER} $\downarrow$ & \textbf{IEMOCAP Acc.} $\uparrow$ & \textbf{MMAU-mini Acc.} $\uparrow$ & \textbf{CoVoST2 BLEU} $\uparrow$ & \textbf{Latency (s)} & \textbf{TFLOPs} \\
\midrule
No Compression & $4.85^{\pm 0.34}$ & $39.34^{\pm 0.29}$ & $54.0^{\pm 0.85}$ & $36.06^{\pm 0.07}$ & $1.15$ & $1.403$ \\
\midrule
\multicolumn{7}{c}{\textit{Compression Ratio 0.36}} \\
\midrule
Random                         & $5.64^{\pm 1.92}$ & $39.00^{\pm 0.51}$ & $52.60^{\pm 0.89}$ & $19.05^{\pm 0.05}$ & $0.97$ & $0.96$ \\
SpeechPrune            & $5.55^{\pm 0.06}$ & $38.28^{\pm 1.10}$ & $53.70^{\pm 0.44}$ & $22.60^{\pm 0.07}$ & $1.01$ & $0.96$ \\
SparseVLM$^{\dagger\ddagger}$ & $5.43^{\pm 0.79}$ & $39.75^{\pm 0.36}$ & $52.70^{\pm 0.43}$ & $22.05^{\pm 0.07}$ & $1.02$ & $1.00$ \\
LTBM                               & $6.60^{\pm 0.00}$ & $40.40^{\pm 0.84}$ & $54.40^{\pm 0.00}$ & $26.76^{\pm 0.19}$ & $1.00$ & $0.96$ \\
\textbf{AudioTETRIS}$^{\dagger}$ & \cellcolor{highlightorange}$4.84^{\pm 1.03}$ & \cellcolor{highlightorange}$40.93^{\pm 0.56}$ & \cellcolor{highlightorange}$55.20^{\pm 0.31}$ & \cellcolor{highlightorange}$31.60^{\pm 0.35}$ & $1.02$ & $1.03$ \\
\midrule
\multicolumn{7}{c}{\textit{Compression Ratio 0.51}} \\
\midrule
Random                         & $7.99^{\pm 0.68}$ & $38.03^{\pm 0.90}$ & $51.30^{\pm 0.64}$ & $16.34^{\pm 0.02}$ & $0.77$ & $0.77$ \\
SpeechPrune            & $5.81^{\pm 2.80}$ & $36.91^{\pm 0.99}$ & $53.20^{\pm 0.92}$ & $13.57^{\pm 0.01}$ & $0.80$ & $0.78$ \\
SparseVLM$^{\dagger\ddagger}$ & $6.53^{\pm 1.13}$ & $39.34^{\pm 0.72}$ & $52.70^{\pm 1.04}$ & $17.92^{\pm 0.08}$ & $0.81$ & $0.84$ \\
LTBM                               & $6.95^{\pm 0.07}$ & $39.74^{\pm 0.95}$ & $53.60^{\pm 0.00}$ & $20.23^{\pm 0.02}$ & $0.79$ & $0.77$ \\
\textbf{AudioTETRIS}$^{\dagger}$ & \cellcolor{highlightorange}$5.78^{\pm 1.49}$ & \cellcolor{highlightorange}$39.90^{\pm 0.20}$ & \cellcolor{highlightorange}$54.10^{\pm 0.32}$ & \cellcolor{highlightorange}$31.13^{\pm 0.30}$ & $0.81$ & $0.88$ \\
\midrule
\multicolumn{7}{c}{\textit{Compression Ratio 0.64}} \\
\midrule
Random                         & $8.49^{\pm 6.06}$ & $35.84^{\pm 0.25}$ & $50.90^{\pm 0.89}$ & $25.29^{\pm 0.02}$ & $0.61$ & $0.62$ \\
SpeechPrune            & $6.38^{\pm 2.01}$ & $35.38^{\pm 0.40}$ & $51.90^{\pm 0.83}$ & $6.21^{\pm 0.03}$  & $0.64$ & $0.63$ \\
SparseVLM$^{\dagger\ddagger}$ & $7.53^{\pm 1.09}$ & $39.04^{\pm 0.56}$ & $52.30^{\pm 0.63}$ & $15.67^{\pm 0.04}$ & $0.65$ & $0.70$ \\
LTBM                               & $6.80^{\pm 0.46}$ & $38.97^{\pm 0.77}$ & $52.10^{\pm 0.45}$ & $15.24^{\pm 0.10}$ & $0.63$ & $0.62$ \\
\textbf{AudioTETRIS}$^{\dagger}$ & \cellcolor{highlightorange}$6.04^{\pm 0.87}$ & \cellcolor{highlightorange}$39.44^{\pm 0.52}$ & \cellcolor{highlightorange}$53.50^{\pm 0.61}$ & \cellcolor{highlightorange}$30.67^{\pm 0.09}$ & $0.67$ & $0.74$ \\
\midrule
\multicolumn{7}{c}{\textit{Compression Ratio 0.75}} \\
\midrule
Random                         & $8.92^{\pm 4.33}$ & $33.54^{\pm 0.34}$ & $50.60^{\pm 1.04}$ & $9.56^{\pm 0.17}$  & $0.40$ & $0.49$ \\
SpeechPrune            & $7.85^{\pm 1.52}$ & $35.91^{\pm 0.48}$ & $51.30^{\pm 0.39}$ & $2.50^{\pm 0.02}$  & $0.46$ & $0.49$ \\
SparseVLM$^{\dagger\ddagger}$ & $8.19^{\pm 2.08}$ & $38.94^{\pm 0.43}$ & $51.80^{\pm 0.48}$ & $13.97^{\pm 0.03}$ & $0.47$ & $0.60$ \\
LTBM                               & $6.89^{\pm 2.34}$ & $38.04^{\pm 0.80}$ & $51.60^{\pm 0.91}$ & $10.80^{\pm 0.09}$ & $0.45$ & $0.49$ \\
\textbf{AudioTETRIS}$^{\dagger}$ & \cellcolor{highlightorange}$6.71^{\pm 2.08}$ & \cellcolor{highlightorange}$39.07^{\pm 0.52}$ & \cellcolor{highlightorange}$52.50^{\pm 0.38}$ & \cellcolor{highlightorange}$30.04^{\pm 0.32}$ & $0.45$ & $0.64$ \\
\bottomrule
\end{tabular}%
}
\begin{tablenotes}
\footnotesize
\item[$\dagger$] In-LM pruning (token elimination performed inside the LLM backbone).
\item[$\ddagger$] Originally developed for the vision modality, adapted here for speech/audio.
\end{tablenotes}
\end{threeparttable}
\end{table}

\begin{table}[t]
\centering
\caption{Extended results for Audio Flamingo 3 across all compression ratios.}
\label{tab:audioflamingo3_appendix}
\begin{threeparttable}
\resizebox{\textwidth}{!}{%
\begin{tabular}{@{}l c c c c c c@{}}
\toprule
\textbf{Method} & \textbf{LibriSpeech WER} $\downarrow$ & \textbf{IEMOCAP Acc.} $\uparrow$ & \textbf{MMAU-mini Acc.} $\uparrow$ & \textbf{CoVoST2 BLEU} $\uparrow$ & \textbf{Latency (s)} & \textbf{TFLOPs} \\
\midrule
No Compression & $3.61^{\pm 0.54}$ & $59.11^{\pm 0.87}$ & $62.09^{\pm 0.45}$ & $37.02^{\pm 0.71}$ & $1.29$ & $7.93$ \\
\midrule
\multicolumn{7}{c}{\textit{Compression Ratio 0.36}} \\
\midrule
Random                         & $4.02^{\pm 1.21}$ & $58.43^{\pm 0.63}$ & $60.97^{\pm 0.23}$ & $36.82^{\pm 0.10}$ & $1.13$ & $6.92$ \\
SpeechPrune            & $4.01^{\pm 2.10}$ & $58.23^{\pm 0.59}$ & $61.42^{\pm 0.83}$ & $36.73^{\pm 0.05}$ & $1.15$ & $7.09$ \\
SparseVLM$^{\dagger\ddagger}$ & $3.78^{\pm 1.45}$ & $57.34^{\pm 0.32}$ & $58.83^{\pm 0.68}$ & $36.18^{\pm 0.06}$ & $1.14$ & $7.31$ \\
LTBM                               & $3.83^{\pm 0.38}$ & $58.11^{\pm 0.41}$ & $61.00^{\pm 0.65}$ & $35.89^{\pm 0.02}$ & $1.21$ & $6.98$ \\
\textbf{AudioTETRIS}$^{\dagger}$ & \cellcolor{highlightorange}$3.71^{\pm 0.45}$ & \cellcolor{highlightorange}$58.94^{\pm 0.59}$ & \cellcolor{highlightorange}$61.77^{\pm 0.37}$ & \cellcolor{highlightorange}$37.31^{\pm 0.19}$ & $1.15$ & $7.34$ \\
\midrule
\multicolumn{7}{c}{\textit{Compression Ratio 0.51}} \\
\midrule
Random                         & $6.67^{\pm 0.91}$ & $56.45^{\pm 0.62}$ & $59.62^{\pm 0.28}$ & $34.72^{\pm 0.15}$ & $0.80$ & $5.61$ \\
SpeechPrune            & $5.18^{\pm 1.49}$ & $57.81^{\pm 0.97}$ & $61.19^{\pm 0.29}$ & $35.18^{\pm 0.16}$ & $0.87$ & $5.71$ \\
SparseVLM$^{\dagger\ddagger}$ & $4.97^{\pm 1.32}$ & $57.39^{\pm 0.49}$ & $58.14^{\pm 0.52}$ & \cellcolor{highlightorange}$36.30^{\pm 0.32}$ & $0.92$ & $6.18$ \\
LTBM                               & $5.87^{\pm 0.18}$ & \cellcolor{highlightorange}$58.91^{\pm 0.56}$ & $60.79^{\pm 0.26}$ & $35.17^{\pm 0.13}$ & $0.91$ & $5.63$ \\
\textbf{AudioTETRIS}$^{\dagger}$ & \cellcolor{highlightorange}$4.62^{\pm 0.97}$ & $58.85^{\pm 0.34}$ & \cellcolor{highlightorange}$61.59^{\pm 0.72}$ & $36.26^{\pm 0.09}$ & $0.85$ & $6.16$ \\
\midrule
\multicolumn{7}{c}{\textit{Compression Ratio 0.64}} \\
\midrule
Random                         & $10.91^{\pm 1.05}$ & $53.29^{\pm 0.33}$ & $58.19^{\pm 0.52}$ & $33.05^{\pm 0.31}$ & $0.73$ & $4.53$ \\
SpeechPrune            & $6.59^{\pm 1.92}$ & $56.71^{\pm 0.63}$ & \cellcolor{highlightorange}$60.48^{\pm 0.64}$ & $34.87^{\pm 0.26}$ & $0.78$ & $4.56$ \\
SparseVLM$^{\dagger\ddagger}$ & $6.08^{\pm 1.94}$ & $53.98^{\pm 0.62}$ & $56.39^{\pm 0.93}$ & $35.39^{\pm 0.09}$ & $0.82$ & $5.14$ \\
LTBM                               & $6.49^{\pm 0.71}$ & $57.18^{\pm 0.21}$ & $59.23^{\pm 0.20}$ & $33.90^{\pm 0.04}$ & $0.83$ & $4.60$ \\
\textbf{AudioTETRIS}$^{\dagger}$ & \cellcolor{highlightorange}$5.72^{\pm 0.63}$ & \cellcolor{highlightorange}$58.31^{\pm 0.26}$ & $59.85^{\pm 0.13}$ & \cellcolor{highlightorange}$35.93^{\pm 0.15}$ & $0.77$ & $5.14$ \\
\midrule
\multicolumn{7}{c}{\textit{Compression Ratio 0.75}} \\
\midrule
Random                         & $15.69^{\pm 1.12}$ & $51.89^{\pm 1.08}$ & $54.58^{\pm 0.42}$ & $27.72^{\pm 0.20}$ & $0.51$ & $3.59$ \\
SpeechPrune            & $11.48^{\pm 2.63}$ & $54.78^{\pm 0.53}$ & $59.59^{\pm 0.97}$ & $31.98^{\pm 0.29}$ & $0.63$ & $3.68$ \\
SparseVLM$^{\dagger\ddagger}$ & $9.59^{\pm 1.27}$ & $52.60^{\pm 0.49}$ & $53.21^{\pm 0.28}$ & $34.25^{\pm 0.02}$ & $0.62$ & $4.25$ \\
LTBM                               & $8.94^{\pm 0.33}$ & $57.31^{\pm 0.29}$ & $57.13^{\pm 0.40}$ & $33.01^{\pm 0.06}$ & $0.66$ & $3.52$ \\
\textbf{AudioTETRIS}$^{\dagger}$ & \cellcolor{highlightorange}$6.39^{\pm 1.03}$ & \cellcolor{highlightorange}$58.09^{\pm 0.37}$ & \cellcolor{highlightorange}$60.92^{\pm 0.29}$ & \cellcolor{highlightorange}$35.94^{\pm 0.14}$ & $0.60$ & $4.26$ \\
\bottomrule
\end{tabular}%
}
\begin{tablenotes}
\footnotesize
\item[$\dagger$] In-LM pruning (token elimination performed inside the LLM backbone).
\item[$\ddagger$] Originally developed for the vision modality, adapted here for speech/audio.
\end{tablenotes}
\end{threeparttable}
\end{table}

\subsubsection{Extended Component-wise Ablation}

\label{sec:appendix_ablation}

Section~\ref{sec:ablation-component} reports the component and stage ablation of AudioTETRIS's fused score $s_t^{(\ell)} = \tilde p_t^{(\ell)} \cdot w_t$ at compression ratio $0.51$ only. Here we provide the extended results across all four evaluated compression ratios ($0.36$, $0.51$, $0.64$, $0.75$) and all four benchmarks (LibriSpeech, IEMOCAP, MMAU-mini, CoVoST2), comparing the isolated single-stage variants (Only Stage 1, Only Stage 2) against the full two-stage AudioTETRIS pipeline (Stage 1 + Stage 2). Table~\ref{tab:ablation_results} shows that combining both stages consistently outperforms either stage in isolation across every compression ratio and benchmark, with the gap widening substantially at higher compression - confirming that the phonetic transition prior and the depth-adaptive attention score contribute complementary, rather than redundant information.

\definecolor{highlightlavender}{HTML}{EAE0F5}
\begin{table}[htbp]
\centering
\caption{Ablation results comparing individual stages against full AudioTETRIS across different compression ratios. The best-performing combination per compression ratio block is highlighted.}
\label{tab:ablation_results}
\resizebox{0.80\linewidth}{!}{%
\begin{tabular}{lcccc}
\toprule
\textbf{Method} & \textbf{LibriSpeech WER} $\downarrow$ & \textbf{IEMOCAP Acc.} $\uparrow$ & \textbf{MMAU-mini Acc.} $\uparrow$ & \textbf{CoVoST2 BLEU} $\uparrow$ \\
\midrule
\multicolumn{5}{c}{\textit{Compression Ratio 0.36}} \\
\midrule
Only Stage 1 & $5.50^{\pm0.25}$ & $57.45^{\pm0.39}$ & $41.80^{\pm0.62}$ & $31.21^{\pm0.09}$ \\
Only Stage 2 & $9.21^{\pm0.36}$ & $57.30^{\pm1.19}$ & $41.70^{\pm0.80}$ & $30.93^{\pm0.04}$ \\
\textbf{Stage 1 + Stage 2} & \cellcolor{highlightlavender}$\mathbf{4.52^{\pm0.21}}$ & \cellcolor{highlightlavender}$\mathbf{58.04^{\pm0.84}}$ & \cellcolor{highlightlavender}$\mathbf{42.36^{\pm0.21}}$ & \cellcolor{highlightlavender}$\mathbf{31.96^{\pm0.04}}$ \\
\midrule
\multicolumn{5}{c}{\textit{Compression Ratio 0.51}} \\
\midrule
Only Stage 1 & $7.39^{\pm0.09}$ & $56.37^{\pm1.55}$ & $41.90^{\pm0.42}$ & $31.18^{\pm0.08}$ \\
Only Stage 2 & $9.42^{\pm0.32}$ & $56.22^{\pm0.21}$ & $41.20^{\pm0.85}$ & $30.55^{\pm0.13}$ \\
\textbf{Stage 1 + Stage 2} & \cellcolor{highlightlavender}$\mathbf{5.77^{\pm0.33}}$ & \cellcolor{highlightlavender}$\mathbf{57.03^{\pm0.41}}$ & \cellcolor{highlightlavender}$\mathbf{42.91^{\pm0.36}}$ & \cellcolor{highlightlavender}$\mathbf{31.45^{\pm0.06}}$ \\
\midrule
\multicolumn{5}{c}{\textit{Compression Ratio 0.64}} \\
\midrule
Only Stage 1 & $8.92^{\pm0.46}$ & $56.68^{\pm1.43}$ & $41.50^{\pm0.14}$ & $30.11^{\pm0.04}$ \\
Only Stage 2 & $9.62^{\pm0.22}$ & $55.77^{\pm0.57}$ & $38.70^{\pm1.13}$ & $30.85^{\pm0.19}$ \\
\textbf{Stage 1 + Stage 2} & \cellcolor{highlightlavender}$\mathbf{6.62^{\pm0.29}}$ & \cellcolor{highlightlavender}$\mathbf{57.90^{\pm0.76}}$ & \cellcolor{highlightlavender}$\mathbf{41.83^{\pm0.15}}$ & \cellcolor{highlightlavender}$\mathbf{31.24^{\pm0.11}}$ \\
\midrule
\multicolumn{5}{c}{\textit{Compression Ratio 0.75}} \\
\midrule
Only Stage 1 & $13.78^{\pm0.23}$ & $55.02^{\pm0.14}$ & $40.80^{\pm0.42}$ & $30.24^{\pm0.09}$ \\
Only Stage 2 & $15.56^{\pm0.57}$ & $50.94^{\pm1.06}$ & $36.10^{\pm0.57}$ & $24.29^{\pm0.11}$ \\
\textbf{Stage 1 + Stage 2} & \cellcolor{highlightlavender}$\mathbf{8.50^{\pm0.18}}$ & \cellcolor{highlightlavender}$\mathbf{55.34^{\pm1.25}}$ & \cellcolor{highlightlavender}$\mathbf{41.46^{\pm0.32}}$ & \cellcolor{highlightlavender}$\mathbf{31.33^{\pm0.07}}$ \\
\bottomrule
\end{tabular}%
}
\end{table}

\subsubsection{Single-Stage and Alternative Multi-Stage Schedules}

\label{app:multi-stage}

Section~\ref{sec:exp-setup} fixes AudioTETRIS's schedule at two stages, layers $\{1,15\}$, uniform per-stage split, with layer 1 placed for compute savings (Appendix~\ref{sec:appendix-early-layer}) and layer 15 set by crossover calibration on Qwen2Audio (Appendix~\ref{sec:appendix-crossover}). Here we ablate two orthogonal design choices: \emph{where} in the decoder the pruning stages are placed and \emph{how} the compression budget is
allocated across stages once placement is fixed.

\textbf{Schedule parameterization.} For stages at layers
$(\ell_1,\dots,\ell_n)$ and target keep-fraction $K = 1-\text{rate}$, we
assign each stage a weight $w_i$ with $\sum_i w_i = 1$ and set the
per-stage retained fraction
\begin{equation}
r_i = K^{w_i}, \qquad \prod_{i=1}^n r_i = K,
\label{eq:multistage-closed-form}
\end{equation}
so every variant reaches the same target compression rate regardless of how $w_i$ is split. Equal weights $w_i = 1/n$ give the uniform split used by the main text; $w_1 > w_2$ front-loads the budget toward the earlier stage, and the reverse back-loads it. Dense variants keep $n$ large and the split uniform, spreading small drops across many stages instead of a few large ones. Table~\ref{tab:multistage-combo-specifics} lists all variants tested. Table~\ref{tab:multistage-rate-specifics} reports the resulting per-stage compression rate $(1-r_i)\times 100\%$ at each evaluated compression rate. For uniform-split variants this is a single number repeated $n$ times; for front-/back-loaded it differs stage to stage, with front-loaded removing a larger share early and back-loaded the reverse.

\begin{table}[h]
\centering
\small
\caption{Candidate pruning schedules. Placement variants use the uniform
split; front-/back-loaded vary only $w_i$ at the default layers $\{1,15\}$.}
\label{tab:multistage-combo-specifics}
\begin{tabular}{l l c c}
\toprule
Schedule & Layers & $n$ stages & Split \\
\midrule
Single-stage @ 1  & 1 & 1 & -- \\
Single-stage @ 15 & 15 & 1 & -- \\
Two-stage $\{1,15\}$ (default) & 1, 15 & 2 & Uniform \\
Two-stage $\{1,19\}$ & 1, 19 & 2 & Uniform \\
Three-stage $\{1,8,15\}$ & 1, 8, 15 & 3 & Uniform \\
Front-loaded & 1, 15 & 2 & $w = (0.7, 0.3)$ \\
Back-loaded & 1, 15 & 2 & $w = (0.3, 0.7)$ \\
Dense (every $\sim$3 layers, full depth) & 1, 4, 7, 10, 13, 16, 19, 22, 25, 28, 31 & 11 & Uniform \\
\bottomrule
\end{tabular}
\end{table}

\begin{table}[h]
\centering
\small
\caption{Per-stage token-drop rate (\% of currently live tokens removed at
that stage) for each schedule, at every evaluated compression rate. Equal
per-stage rates are written once as ``$X\%$ ($\times n$)''; front-/back-loaded
list rates stage by stage.}
\label{tab:multistage-rate-specifics}
\resizebox{\linewidth}{!}{%
\begin{tabular}{l l c c c c}
\toprule
Schedule & Layers & rate 0.36 & rate 0.51 & rate 0.64 & rate 0.75 \\
\midrule
Single-stage @ 1  & 1      & 36\% & 51\% & 64\% & 75\% \\
Single-stage @ 15 & 15     & 36\% & 51\% & 64\% & 75\% \\
Two-stage (default) & 1, 15  & 20\% ($\times$2) & 30\% ($\times$2) & 40\% ($\times$2) & 50\% ($\times$2) \\
Two-stage $\{1,19\}$ & 1, 19  & 20\% ($\times$2) & 30\% ($\times$2) & 40\% ($\times$2) & 50\% ($\times$2) \\
Three-stage & 1, 8, 15 & 14\% ($\times$3) & 21\% ($\times$3) & 29\% ($\times$3) & 37\% ($\times$3) \\
Front-loaded & 1, 15 & 27\%/13\% & 39\%/19\% & 51\%/26\% & 62\%/34\% \\
Back-loaded & 1, 15 & 13\%/27\% & 19\%/39\% & 26\%/51\% & 34\%/62\% \\
Dense (11 stages) & 1,4,\dots,31 & 4\% ($\times$11) & 6\% ($\times$11) & 9\% ($\times$11) & 12\% ($\times$11) \\
\bottomrule
\end{tabular}%
}
\end{table}

\begin{table}[h]
\centering
\small
\caption{Schedule ablation, LibriSpeech WER (\%, lower is better), Qwen2-Audio, across compression rates.}
\label{tab:multistage-wer-results}
\begin{tabular}{l c c c c}
\toprule
Schedule & rate 0.36 & rate 0.51 & rate 0.64 & rate 0.75 \\
\midrule
Single-stage @ 1               & $6.51^{\pm 0.14}$          & $8.92^{\pm 0.17}$          & $12.40^{\pm 0.36}$          & $18.78^{\pm 0.41}$ \\
Single-stage @ 15              & $5.44^{\pm 0.61}$          & $6.93^{\pm 0.42}$          & $8.94^{\pm 0.48}$           & $12.89^{\pm 0.82}$ \\
Two-stage $\{1,15\}$ (default) & $4.52^{\pm 0.21}$          & $\mathbf{5.77^{\pm 0.33}}$ & $6.62^{\pm 0.29}$           & $8.50^{\pm 0.18}$  \\
Two-stage $\{1,19\}$           & $4.78^{\pm 0.38}$          & $6.02^{\pm 0.35}$          & $6.94^{\pm 0.62}$           & $8.96^{\pm 0.31}$  \\
Three-stage $\{1,8,15\}$       & $\mathbf{4.36^{\pm 0.71}}$ & $5.93^{\pm 0.71}$          & $\mathbf{6.47^{\pm 0.72}}$  & $\mathbf{8.12^{\pm 0.73}}$ \\
Front-loaded                   & $4.99^{\pm 0.69}$          & $6.72^{\pm 0.76}$          & $7.68^{\pm 0.29}$           & $10.73^{\pm 0.28}$ \\
Back-loaded                    & $5.73^{\pm 0.31}$          & $7.34^{\pm 0.62}$          & $9.38^{\pm 0.34}$           & $14.53^{\pm 0.42}$ \\
Dense (11 stages)              & $7.13^{\pm 0.45}$          & $8.79^{\pm 0.57}$          & $10.09^{\pm 0.62}$          & $12.92^{\pm 0.36}$ \\
\bottomrule
\end{tabular}
\end{table}

\begin{table}[h]
\centering
\small
\caption{Efficiency and downstream-task trade-offs at compression rate 0.51.
Latency is relative to the uncompressed baseline ($1.00\times$).}
\label{tab:multistage-tradeoffs}
\begin{tabular}{l c c c c c c}
\toprule
Schedule & WER & IEMOCAP & MMAU & CoVoST & Latency & TFLOPs \\
\midrule
Single-stage @ 1               & $8.92^{\pm 0.17}$          & $53.55^{\pm 0.50}$          & $39.81^{\pm 0.41}$          & $27.05^{\pm 0.05}$          & $\mathbf{0.98}$ & $\mathbf{5.51}$ \\
Single-stage @ 15              & $6.93^{\pm 0.42}$          & $55.83^{\pm 0.52}$          & $41.52^{\pm 0.24}$          & $29.53^{\pm 0.09}$          & $0.99$          & $5.92$ \\
Two-stage $\{1,15\}$ (default) & $\mathbf{5.77^{\pm 0.33}}$ & $57.03^{\pm 0.41}$          & $42.91^{\pm 0.36}$          & $31.45^{\pm 0.06}$          & $1.00$          & $6.13$ \\
Two-stage $\{1,19\}$           & $6.02^{\pm 0.35}$          & $57.41^{\pm 0.47}$          & $43.18^{\pm 0.18}$          & $\mathbf{31.69^{\pm 0.07}}$ & $1.03$          & $6.13$ \\
Three-stage $\{1,8,15\}$       & $5.93^{\pm 0.71}$          & $55.68^{\pm 0.37}$          & $41.67^{\pm 0.72}$          & $31.45^{\pm 0.10}$          & $1.04$          & $6.32$ \\
Front-loaded                   & $6.72^{\pm 0.76}$          & $56.91^{\pm 0.19}$          & $41.48^{\pm 0.39}$          & $30.20^{\pm 0.18}$          & $1.02$          & $5.71$ \\
Back-loaded                    & $7.34^{\pm 0.62}$          & $55.93^{\pm 0.31}$          & $41.44^{\pm 0.53}$          & $26.68^{\pm 0.04}$          & $1.06$          & $6.21$ \\
Dense (11 stages)              & $8.79^{\pm 0.57}$          & $\mathbf{57.62^{\pm 0.81}}$ & $\mathbf{43.30^{\pm 0.64}}$ & $30.90^{\pm 0.07}$          & $1.21$          & $6.93$ \\
\bottomrule
\end{tabular}
\end{table}

\textbf{Results.} We find that single-stage compression at layer 15 (6.93\% WER at rate 0.51; Table~\ref{tab:multistage-wer-results}) consistently outperforms layer 1 (8.92\%), highlighting the role of text-conditioned attention in token ranking. Splitting the compression budget across two stages improves upon single-stage baselines across all rates, regardless of the chosen layers. Within two-stage setups, our default $\{1,15\}$ schedule achieves the strongest performance (5.77\% at rate 0.51), outperforming $\{1,19\}$ (6.02\%) by 0.2-0.5 percentage points across the sweep.

The three-stage schedule $\{1,8,15\}$ yields small gains over the two-stage default at three of the four evaluated rates - $4.36\%$ vs.\ $4.52\%$ WER at rate $0.36$, $6.47\%$ vs.\ $6.62\%$ at rate $0.64$, and $8.12\%$ vs.\ $8.50\%$ at rate $0.75$ - but this advantage is not monotonic: at rate $0.51$ the ordering reverses, with the two-stage default outperforming three-stage ($5.77\%$ vs.\ $5.93\%$). Given this inconsistency and the added latency overhead of a third pruning stage (Table~\ref{tab:multistage-tradeoffs}), we retain the two-stage $\{1,15\}$ schedule as the default configuration used throughout the main text, rather than the marginally stronger but less consistent three-stage variant.

Redistributing the budget within two-stage setups reveals a clear asymmetry: front-loading (larger early cut) remains competitive with the default (4.99\% vs.\ 4.52\% at rate 0.36), whereas back-loading degrades sharply under higher compression (14.53\% vs.\ 10.73\% at rate 0.75). This aligns with our analysis in Appendix~\ref{sec:appendix-crossover}: heavy pruning in later layers, where cross-modal attention dominates, incurs a higher penalty than pruning early while representations remain largely intra-audio. Finally, the 11-stage dense schedule underperforms at all rates, showing that frequent, fine-grained pruning could accumulate more errors for the ASR task.

Table~\ref{tab:multistage-tradeoffs} highlights the efficiency trade-offs at rate 0.51. We observe that the two-stage compression schedule offers one of the most favorable balance between accuracy and computational cost. In contrast, three-stage and dense schedules introduce a slight overhead ($1.04\times$ and $1.21\times$ latency, respectively) without consistent downstream gains, confirming the 2-stage compression is an effective overall configuration.

\subsubsection{Qualitative Examples of Token Survival}

\label{app:token_survival_viz}
\begin{figure}[t]
    \centering
    \begin{subfigure}[t]{\linewidth}
        \includegraphics[width=0.85\linewidth]{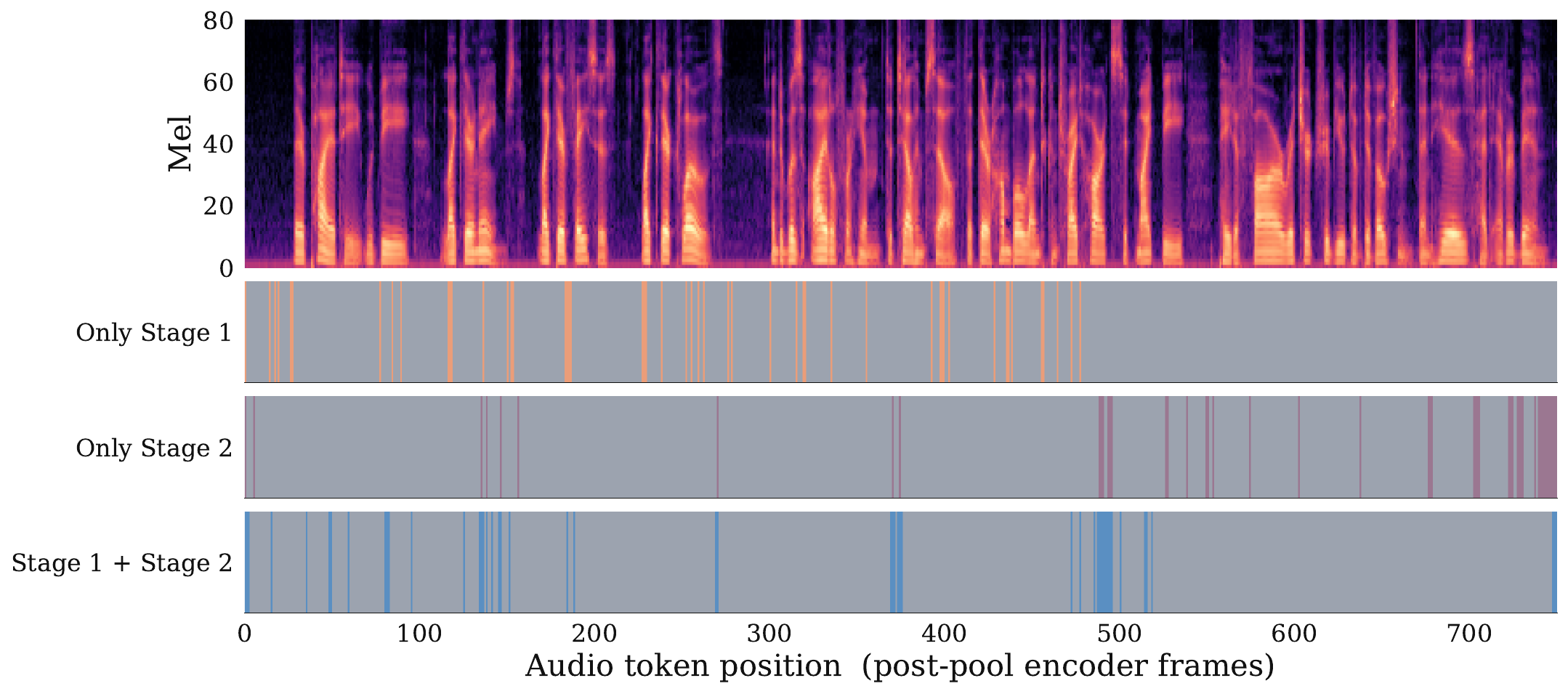}
        \caption{Example~1: \textit{``their piety would be like their names, like their faces, like their clothes\ldots a gift tenfold more acceptable than his elaborate adoration''}}
        \label{fig:token-survival-b}
    \end{subfigure}
    \vspace{0.8em}
    \begin{subfigure}[t]{\linewidth}
        \includegraphics[width=0.85\linewidth]{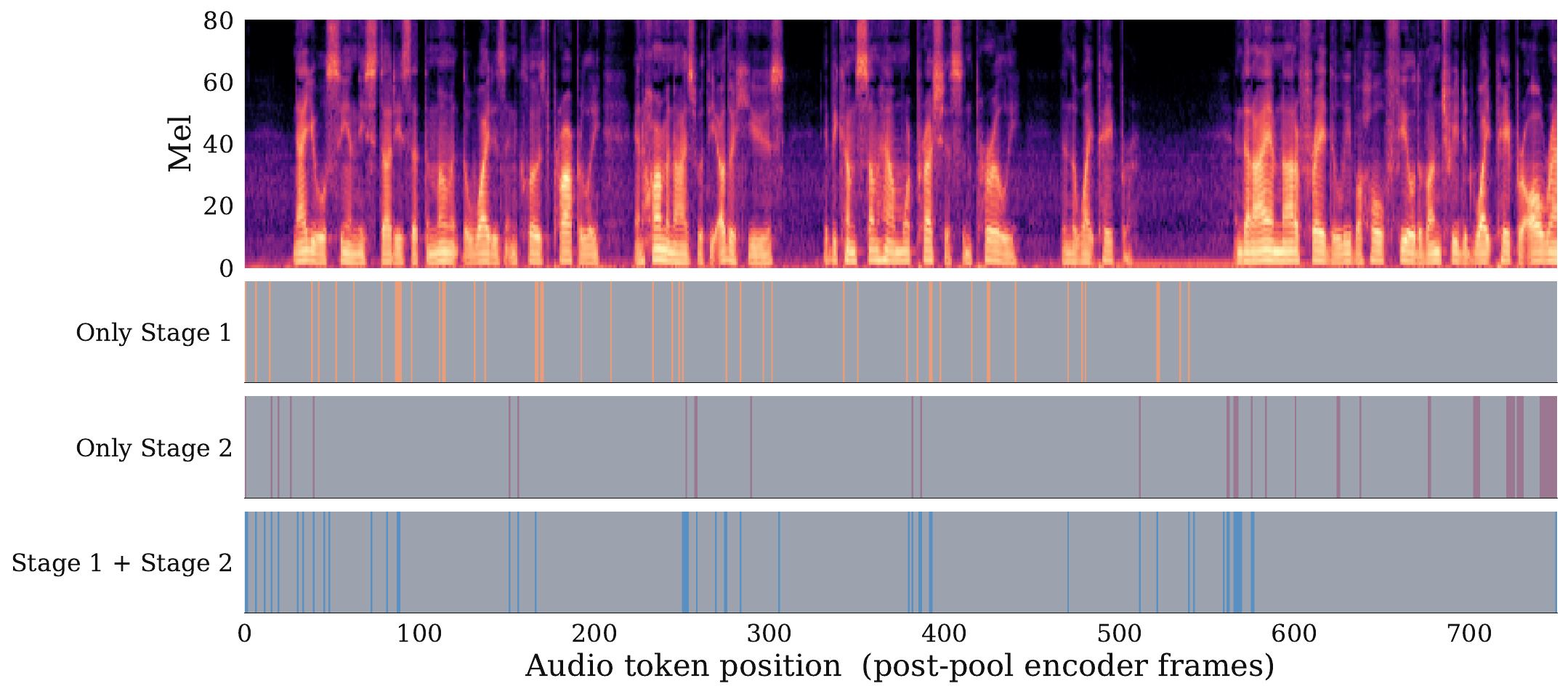}
        \caption{Example~2: \textit{``now you will see in these studies that the moment the white is inclosed properly\ldots the little diamonds in the round window will tell as jewels if they are gradated justly''}}
        \label{fig:token-survival-c}
    \end{subfigure}
    \vspace{0.8em}
    \begin{subfigure}[t]{\linewidth}
        \includegraphics[width=0.85\linewidth]{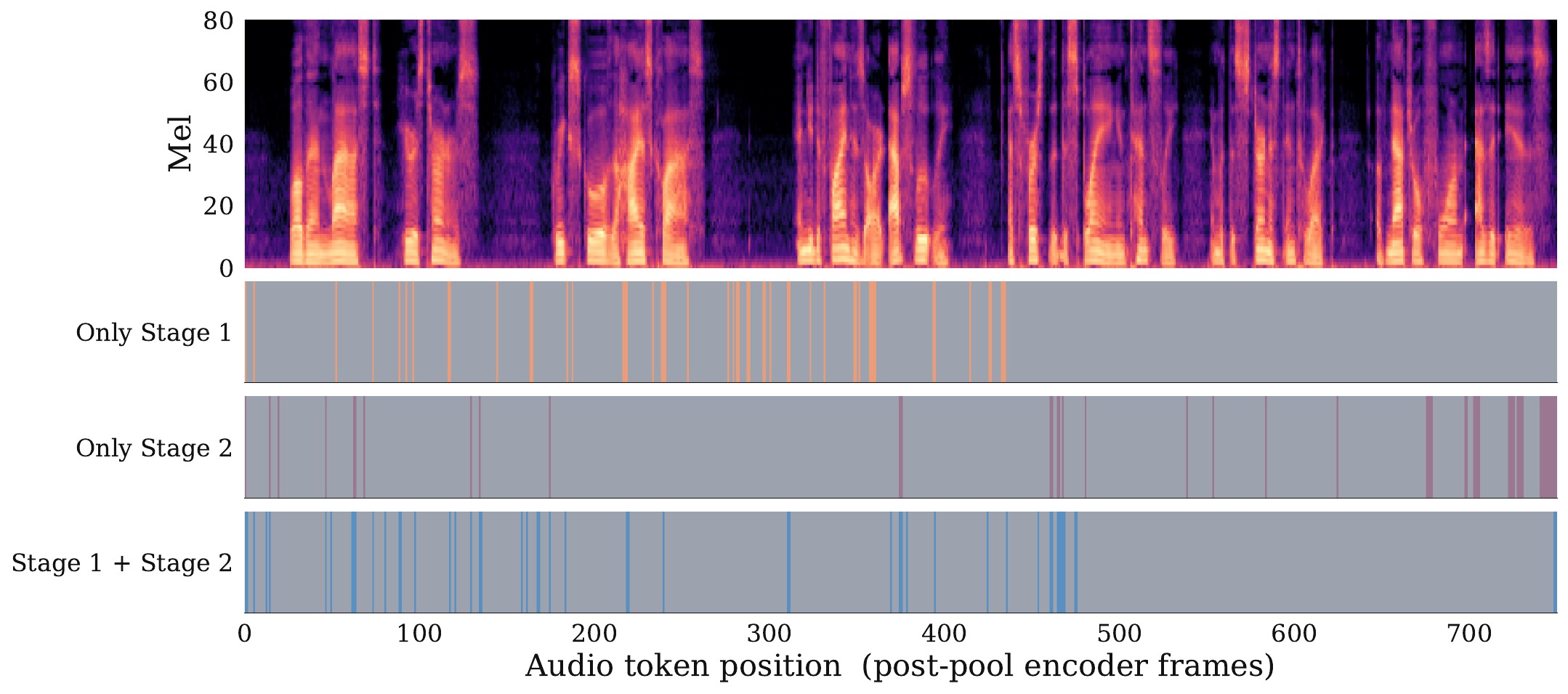}
        \caption{Example~3: \textit{``well then, last, here is turner's greek school of the highest class, and you define his art absolutely as first the displaying intensely and with the sternest intellect of natural form as it is, and then the envelopment of it with cloud and fire''}}
        \label{fig:token-survival-d}
    \end{subfigure}
    \caption{Token survival at the final pruning layer for three representative utterances (ground-truth transcripts shown under each panel). Rows within each panel (top to bottom): mel
    spectrogram, Only Stage~1 (phoneme-transitions only), Only Stage~2 (attention-only), Stage 1 + Stage 2 (attention $\times$ phoneme-transitions, our full method). Orange = kept uniquely by Only Stage~1; mauve = kept uniquely by Only Stage~2; blue = kept uniquely by Stage 1 + Stage 2; grey = pruned.}
    \label{fig:token-survival-appendix}
\end{figure}

Figure~\ref{fig:token-survival-appendix} shows token survival at the final pruning layer for three representative LibriSpeech utterances. The examples are consistent with our quantitative finding (\S\ref{sec:ablation}) that the two stages contribute complementary rather than redundant information: Stage~1 supplies dense, distributed protection against losing phonetic detail, while Stage~2 contributes a smaller number of additional, task-relevant tokens that Stage~1 alone would not have prioritized.

\end{document}